\documentclass[floats,floatfix,showpacs,amssymb,prd,twocolumn,superscriptaddress,nofootinbib,reprint,longbibliography]{revtex4-1}

\usepackage{amssymb,amsmath,verbatim,mathtools,needspace,enumitem,etoolbox,graphicx,physics,microtype,afterpage,gensymb,tabularx}
\usepackage[dvipsnames, usenames]{xcolor}

\definecolor{linkcolor}{rgb}{0.0,0.3,0.5}
\usepackage[unicode, colorlinks=true, linkcolor=linkcolor, citecolor=linkcolor, filecolor=linkcolor,urlcolor=linkcolor, pdfusetitle]{hyperref}
\usepackage[all]{hypcap}
\usepackage[T1]{fontenc}
\usepackage[utf8]{inputenc}
\usepackage[normalem]{ulem}
\usepackage{bm}

\newcommand{\ssim}{\mathchar"5218\relax\,}

\newcommand\orcid[1]{\href{https://orcid.org/#1}{$\!\!$\includegraphics[scale=0.006]{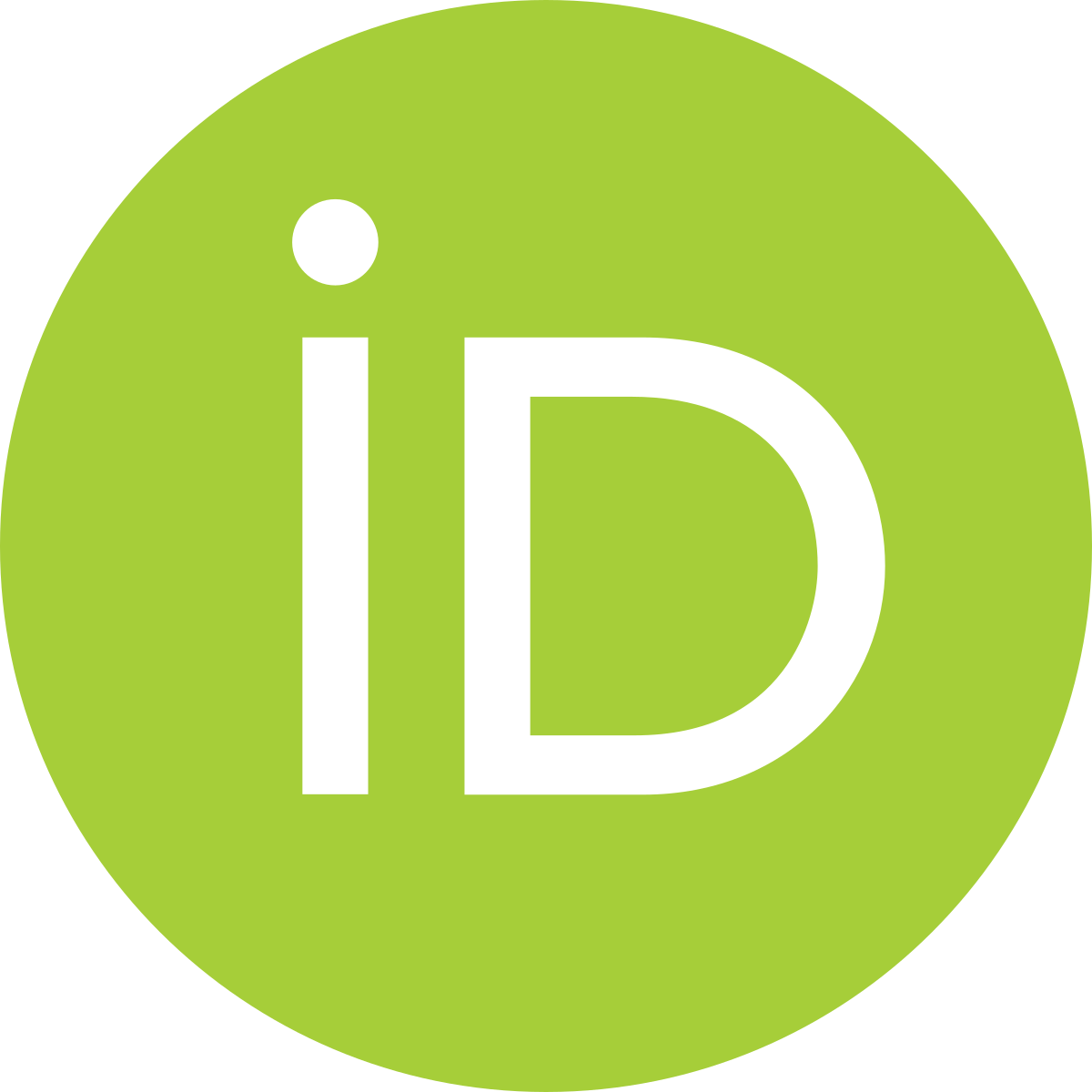} $\!\!$}}

\makeatletter
\newcommand*{\balancecolsandclearpage}{\close@column@grid \cleardoublepage \twocolumngrid}
\makeatother

\graphicspath{%
  {figs/}%
}

\newcommand{\bham}{\affiliation{School of Physics and Astronomy and Institute for Gravitational Wave Astronomy, \\ University of Birmingham, Birmingham, B15 2TT, United Kingdom}}
\newcommand{\cornell}{\affiliation{Cornell Center for Astrophysics
    and Planetary Science, Cornell University, Ithaca, New York 14853, USA}}
\newcommand\cornellPhys{\affiliation{Department of Physics, Cornell
    University, Ithaca, New York 14853, USA}}
\newcommand{\caltech}{\affiliation{TAPIR 350-17, California Institute of
Technology, 1200 East California Boulevard, Pasadena, California 91125, USA}}
\newcommand{\AEI}{\affiliation{Max Planck Institute for Gravitational
    Physics (Albert Einstein Institute), Am M\"uhlenberg 1, Potsdam 14476,
    Germany}} %

\DeclareMathAlphabet{\mathpzc}{OT1}{pzc}{m}{it}

\newcommand{\h}{\mathpzc{h}}
\newcommand{\hlm}{\mathpzc{h}_{\ell m}}

\newcommand{\torb}{t_\mathrm{orb}}
\newcommand{\tpre}{t_\mathrm{pre}}

\newcommand{\Norb}{N_\mathrm{orb}}
\newcommand{\Npre}{N_\mathrm{pre}}
\newcommand{\rudp}{r_{\mathrm{ud}+}}
\newcommand{\rudm}{r_{\mathrm{ud}-}}
\newcommand{\rudpm}{r_{\mathrm{ud}\pm}}

\newcommand{\bchi}{\bm{\chi}}
\newcommand{\thetapert}{\theta_\mathrm{pert}}
\newcommand{\thetaal}{\theta_i^{\mathrm{align}}}
\newcommand{\thetaL}{\theta_L}
\newcommand{\dtheta}{\Delta\theta}

\newcommand{\tmerger}{t_\mathrm{merger}}
\newcommand{\omegaorb}{\omega_\mathrm{orb}}

\newcommand{\zhat}{\widehat{\bm{z}}}
\newcommand{\Lhat}{\widehat{\bm{L}}}

\newcommand{\chihat}{\widehat{\bm{\chi}}}

\renewcommand{\deg}{^\circ}

\begin{document}

\title{Up-down instability of binary black holes  in numerical relativity}

\author{Vijay Varma \orcid{0000-0002-9994-1761}}
\email{vvarma@cornell.edu}
\thanks{Klarman fellow}
\cornellPhys
\cornell
\caltech

\author{Matthew Mould \orcid{0000-0001-5460-2910}}
\bham

\author{Davide Gerosa \orcid{0000-0002-0933-3579}}
\bham

\author{\\Mark A. Scheel \orcid{0000-0001-6656-9134}}
\caltech

\author{Lawrence E. Kidder \orcid{0000-0001-5392-7342}}
\cornell

\author{Harald P. Pfeiffer \orcid{0000-0001-9288-519X}}
\AEI

\hypersetup{pdfauthor={Varma et al.}}

\date{\today}

\begin{abstract}
Binary black holes with spins that are aligned with the orbital angular
momentum do not precess. However, post-Newtonian calculations predict that
``up-down'' binaries, in which the spin of the heavier (lighter) black hole is
aligned (antialigned) with the orbital angular momentum, are unstable when the
spins are slightly perturbed from perfect alignment. This instability provides
a possible mechanism for the formation of precessing binaries in environments
where sources are preferentially formed with (anti) aligned spins. In this
paper, we present the first full numerical relativity simulations capturing
this instability.  These simulations span $\sim 100$ orbits and $\sim 3$--$5$
precession cycles before merger, making them some of the longest numerical
relativity simulations to date.  Initialized with a small perturbation of
$1^{\circ}$--$10^{\circ}$, the instability causes a dramatic growth of the spin
misalignments, which can reach $\sim 90^{\circ}$ near merger. We show that this
leaves a strong imprint on the subdominant modes of the gravitational wave
signal, which can potentially be used to distinguish up-down binaries from
other sources. Finally, we show that post-Newtonian and effective-one-body
approximants are able to reproduce the unstable dynamics of up-down binaries
extracted from numerical relativity.
\end{abstract}

\maketitle

\section{Introduction}
\label{sec:introduction}

The detections of gravitational waves (GWs) emitted by the inspiral and merger
of stellar-mass black hole (BH) binaries are now regular events for LIGO and
Virgo~\cite{LIGOScientific:2018mvr, Abbott:2020niy}.
Upgrades to detector sensitivities and waveform models, and increasing catalog
size will lead to improved inference on the parameters of individual detections
as well as those of the underlying source
population~\cite{LIGOScientific:2018jsj, Abbott:2020gyp, Roulet:2020wyq}. Of
particular importance are the BH spins, which for generic binaries
are tilted with respect to the orbital angular momentum and can cause
significant modulations in the emitted GW signal due to precession of the
orbital plane~\cite{Apostolatos:1994pre, Kidder:1995zr}.  Spin orientations are
powerful observables for determining the astrophysical formation channels of GW
events~\cite{Gerosa:2013laa, Rodriguez:2016vmx, Vitale:2015tea, Talbot:2017yur,
Farr:2017uvj, Stevenson:2017dlk, Gerosa:2018wbw, Belczynski:2017gds}.

Though in general the BH spins will change direction over the inspiral,
configurations in which both spins are aligned with the orbital angular
momentum of the binary are equilibrium solutions of the spin precession
problem. Due to their regular behavior and simpler dynamics, BH binaries with
aligned spins have been used extensively to construct waveform models,
implement GW searches, and perform numerical-relativity (NR) simulations. They
are also interesting from an astrophysical standpoint, since stellar-mass BH
binary formation via isolated stellar evolution~\cite{Kalogera:1999tq,
Gerosa:2018wbw} or embedment in gaseous disks~\cite{Yang:2019okq,
McKernan:2019beu} may lead to BH binaries with small spin tilts. On the
contrary, large misalignment are expected for binaries formed in cluster
environments~\cite{Mandel:2009nx}. For supermassive BHs targeted by the LISA
mission, spin orientations might help distinguishing between gas-rich and
gas-poor host galaxies~\cite{Miller:2013gya, Sesana:2014bea, Gerosa:2015xya}.

For unequal-mass systems, there are four distinct aligned-spin configurations.
Referring to the direction of a component BH spin that is aligned
(antialigned) with the orbital angular momentum as ``up'' (``down''), the four
alignments are up-up, down-down, down-up and up-down. In this notation, the
direction before (after) the hyphen labels spin of the heavier (lighter) BH. As
first pointed out by \citeauthor{Gerosa:2015hba}~\cite{Gerosa:2015hba}, only
the former three configurations are stable equilibria. On the other hand,
up-down binaries, where the spin of the heavier (lighter) BH is aligned
(antialigned) with the orbital angular momentum, can become unstable (see also
Refs.~\cite{Lousto:2016nlp, Mould:2020cgc}). More specifically, for a binary BH
with component masses $m_1\geq m_2$, total mass $M=m_1+m_2$, mass ratio
$q=m_2/m_1\leq1$, and dimensionless component spin magnitudes $\chi_1$ and $\chi_2$
[where index 1 (2) corresponds to the heavier (lighter) BH], there exist
critical orbital separations (in geometrical units $G=c=1$)
\begin{align}
\rudpm
=\frac{\left(\sqrt{\chi_1} \pm \sqrt{q\chi_2}\right)^4}{\left(1-q\right)^2} M\,,
\end{align}
such that the up-down configuration is unstable for orbital separations $r$ in
the range $\rudp>r>\rudm$.

An up-down binary BH that forms at a large separation $r>\rudp$ with
(infinitesimally small) perturbations to the spin directions remains near its
initial configuration until reaching $r=\rudp$. Upon inspiraling past this
threshold, the binary becomes unstable and begins to precess, leading to large
tilts between the BH spins and the orbital angular momentum. Alternatively, a
perturbed up-down binary initialized within the range $\rudp>r>\rudm$ will be
immediately unstable, while one initialized at $r<\rudm$ will be stable (note,
however, that $\rudm \lesssim M$ for most binary parameters).

This effect was further investigated in Ref.~\cite{Mould:2020cgc}, which showed
that up-down binaries inspiraling from large separation evolve toward specific,
predictable spin configurations after hitting the instability onset. The
up-down instability therefore provides the means by which binary BH spins
initially (anti) aligned by astrophysical formation can become misaligned and
precessing in the sensitivity window of ground- and space-based GW
interferometers.  This may be the case, e.g., for stellar-mass BHs which are
captured by and subsequently merge within the accretion disk of an active
supermassive BH~\cite{Bellovary:2015ifg, Bartos:2016dgn, Stone:2016wzz,
McKernan:2017umu, Yang:2019okq, Yang:2019cbr, McKernan:2019beu}.

But this is only true if such unstable behavior persists until merger. Both the
occurrence~\cite{Gerosa:2015hba} and the end point~\cite{Mould:2020cgc} of the
up-down instability were derived using the multitimescale post-Newtonian (PN)
framework of Refs.~\cite{Kesden:2014sla,Gerosa:2015tea}. While PN techniques
accurately describe the binary dynamics during the earlier inspiral, they
inevitably fail to capture strong-field effects near merger. The only accurate
solutions to the full general relativistic two-body problem are currently
provided by NR simulations.

\begin{center}
\begin{table}
\begin{tabular}{|c|c |c |c |c |c |c |c |}
\hline
ID & $~~q~~$ & $\chi_{1, 2}$ & $\theta_{\rm pert}$ & Config.  & $M\omegaorb$ & $N_{\rm orb}$ & $\tmerger/M$ \\ [0.5ex]
\hline\hline
2313 & 0.9 & 0.8 & $1^{\circ}$ & up-up & 0.0058 & 106 & 68044 \\
\hline
2314 & 0.9 & 0.8 & $1^{\circ}$ & down-down & 0.0055 & 94 & 67681 \\
\hline
2315 & 0.9 & 0.8 & $1^{\circ}$ & down-up & 0.0057 & 100 & 67965 \\
\hline
2316 & 0.9 & 0.8 & $1^{\circ}$ & up-down & 0.0057 & 100 & 67973 \\
\hline
2317 & 0.9 & 0.8 & $5^{\circ}$ & up-up & 0.0058 & 106 & 68005 \\
\hline
2318 & 0.9 & 0.8 & $5^{\circ}$ & down-down & 0.0055 & 94 & 67679 \\
\hline
2319 & 0.9 & 0.8 & $5^{\circ}$ & down-up & 0.0057 & 100 & 67978 \\
\hline
2320 & 0.9 & 0.8 & $5^{\circ}$ & up-down & 0.0058 & 96 & 63924 \\
\hline
2321 & 0.9 & 0.8 & $10^{\circ}$ & up-up & 0.0058 & 106 & 67970 \\
\hline
2322 & 0.9 & 0.8 & $10^{\circ}$ & down-down & 0.0055 & 94 & 67685 \\
\hline
2323 & 0.9 & 0.8 & $10^{\circ}$ & down-up & 0.0057 & 100 & 67943 \\
\hline
2324 & 0.9 & 0.8 & $10^{\circ}$ & up-down & 0.0058 & 96 & 63926 \\
\hline
\end{tabular}
\caption{
The parameters of the 12 NR simulations performed in this study.  We provide
the identifier for each simulation within the SXS catalog~\cite{SXSWebsite}.
Each simulation has mass ratio $q=0.9$ and spin magnitudes
$\chi_{1}=\chi_{2}=0.8$. For each of the four spin configurations (up-up,
down-down, down-up and up-down) we perform three simulations with initial
spin misalignments $\thetapert=1\deg,5\deg,10\deg$ with respect to the perfectly
aligned-spin configuration. The initial orbital frequency $\omegaorb$ is chosen
such that the initial separation is $r=30M$. We also report the merger time
$t_{\rm merger}$ and number of orbits $N_{\rm orb}$.
}
\label{tab:NRdata}
\end{table}
\end{center}

NR simulations of the up-down instability have so far been elusive because of
their high computational cost.  The instability is a precessional effect, and
its observation thus requires a binary to complete at least one, and ideally
several, precession cycles. This results in very long simulations because spin
precession happens on a longer timescale, $t_{\rm pre}/M \sim
(r/M)^{5/2}(1+q)^2/q$, compared to the orbital period, $t_{\rm orb}/M \sim
(r/M)^{3/2}$. Sampling $\Npre$ precession cycles requires simulating a number
of orbits $\Norb = \Npre\tpre/\torb \sim \Npre (r/M)(1+q)^2/q$. In
Ref.~\cite{Mould:2020cgc} it was observed that the precessional instability
appreciably develops over a typical decrease $\sim25M$ in the orbital
separation, resulting in simulations with $\mathcal{O}(100)$ orbits. For
context, typical NR simulations cover $\lesssim20$ orbits~\cite{Boyle:2019kee},
while the longest numerical relativity simulation performed to date covers 175
orbits~\cite{Szilagyi:2015rwa}. While the astrophysically relevant scenario, in
which an up-down binary is initialized in the stability regime ($r>\rudp$) and
becomes unstable, remains prohibitive, simulations instead initialized within
the instability regime ($\rudp>r>\rudm$) are still challenging but possible
with current capabilities.

In this paper, we present the first NR simulations of unstable up-down binaries
and confirm that earlier PN predictions hold in the highly dynamical,
strong-field regime of general relativity (GR). The rest of the paper proceeds as follows. In
Sec.~\ref{sec:nr}, we describe our NR runs. In Sec.~\ref{sec:results}, we
present our results. In particular, we (i) observe the precessional instability
in the up-down simulations, (ii) compare the stable and unstable
configurations, and (iii)  compare NR results against PN and effective-one-body
(EOB) predictions. In Sec.~\ref{sec:conclusions}, we present our conclusions.

\begin{figure*}[p]
\centering
\includegraphics[width=0.85\linewidth]{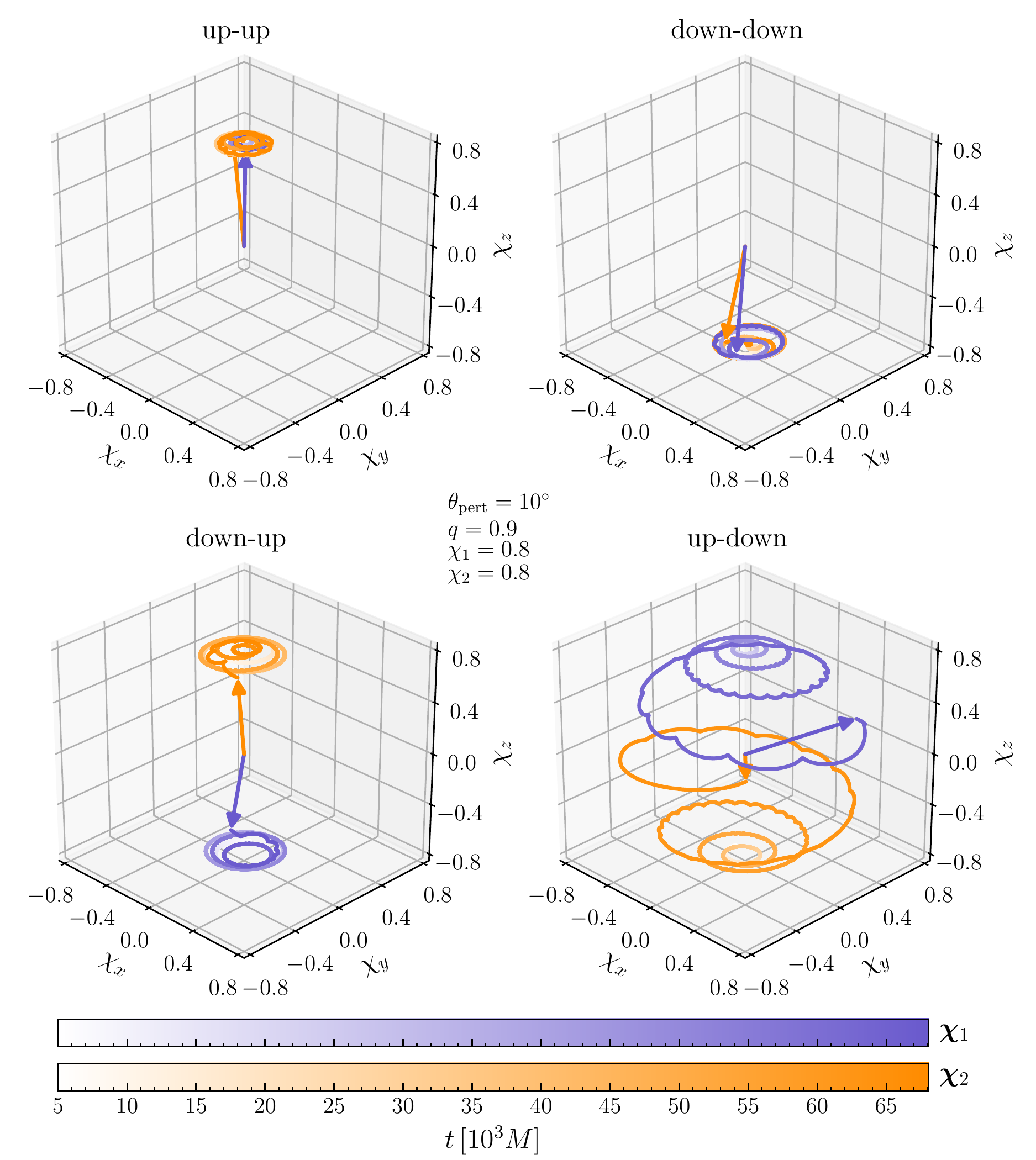}
\caption{
The up-down instability is demonstrated in NR. We consider binaries with mass
ratios $q=0.9$ and spin magnitudes $\chi_{1}=\chi_{2}=0.8$, while the component
spin vectors are initially perturbed from a perfectly aligned-spin
configuration by an angle $\thetapert=10^\circ$. The purple (orange) arrows
represent the spin $\bm{\chi}_1$ ($\bm{\chi}_2$) of the heavier (lighter) BH
near merger.  The colored curves trace the evolution of $\bm{\chi}_{1,2}(t)$ as
the binary precesses. Colors darken linearly in time as indicated on the
color bars. The spins of the up-up (top-left), down-down (top-right), and
down-up (bottom-left) configurations precess stably about $\zhat$, while those
of the up-down (bottom-right) configuration are unstable and become largely
misaligned.  Smaller modulations occur on the shorter orbital timescale.
An animated version of this figure is available at \href{https://davidegerosa.com/spinprecession}{www.davidegerosa.com/spinprecession}.
}
\label{fig:3d}
\end{figure*}

\section{Numerical relativity simulations}
\label{sec:nr}

NR simulations for this work are performed using the Spectral Einstein Code
(SpEC)~\cite{Boyle:2019kee} developed by the Simulating eXtreme Spacetimes
(SXS) Collaboration~\cite{SXSWebsite}. We perform 12 new NR simulations that
have been assigned the identifiers SXS:BBH:2313--2324 and are made publicly
available through the SXS catalog.  The parameters of our runs are summarized
in Table~\ref{tab:NRdata}. For each simulation, we take mass ratio $q=0.9$ and
dimensionless spins $\chi_1=\chi_2=0.8$. This is because the up-down
instability effect is most prominent for binaries with close to (but not
identically) equal masses and large spin magnitudes.

The initial binary separation is chosen to be $r=30M$, which is much smaller
than the instability threshold PN prediction $\rudp\simeq900M$ (note also that
$\rudm\ssim10^{-4}M$). Ideally, we would like to initialize the binary at a
separation $r>900M$ so we can observe an initially stable binary develop the
instability as it passes through $r=\rudp$.  While this would be more
astrophysically realistic, unfortunately, $r\ssim900M$ would lead to
$\ssim5\!\times\!10^{5}$ orbits before merger, which is well outside the
capability of current NR codes. Choosing $r=30M$ as the initial separation
ensures that the instability has sufficient time to develop: these up-down
binaries should already be unstable at their initial separation while at the
same time undergoing several precession cycles before merger (cf.
Fig.~\ref{fig:3d}). Our simulations include $\simeq100$ orbits, $\simeq3$--$5$
precession cycles, and cover a time $\tmerger\simeq65\times10^3M$, making them
some of the longest simulations in the SXS catalog~\cite{Boyle:2019kee}, with
each simulation requiring about $10^5$ CPU hours.

Besides the up-down cases, we perform control simulations in the other three
aligned-spin configurations (up-up, down-down, and down-up), which are predicted
to be stable. We introduce an initial perturbation to the perfectly aligned
configuration to seed the instability. In particular, we consider three
initial values $\thetapert=1\deg,5\deg,10\deg$ of the angles between the BH
spins and the orbital angular momentum direction. The angle between the
in-plane spin components and the separation vector going from the lighter to
the heavier BH are arbitrarily set to $\Phi_1=0$ and $\Phi_2=\pi/2$. The
precise values of $\Phi_1$ and $\Phi_2$ have a negligible impact because these
simulations span several precession cycles.

The waveform is extracted at several extraction spheres at varying finite radii
from the origin and then extrapolated to future null
infinity~\cite{Boyle:2019kee, Boyle:2009vi}. The extrapolated waveforms are
then corrected to account for the initial drift of the center of
mass~\cite{Boyle:2015nqa}. We denote the waveform modes at future null
infinity, scaled to unit mass and unit distance, as $\hlm(t)$. These enter the
complex strain
\begin{equation}
    \h(t, \iota, \varphi) = \sum^{\infty}_{\ell=2} \sum_{m=-l}^{l}
        \hlm(t) ~_{-2}Y_{\ell m}(\iota, \varphi)\,,
\label{eq:spherical_harm}
\end{equation}
where $_{-2}Y_{\ell m}$ are the spin$\,=\!\!-2$ weighted spherical harmonics,
and $\iota$ and $\varphi$ are the polar and azimuthal angles, respectively, on the sky in the
source frame.

\begin{figure*}[thb]
\centering
\includegraphics[width=\linewidth]{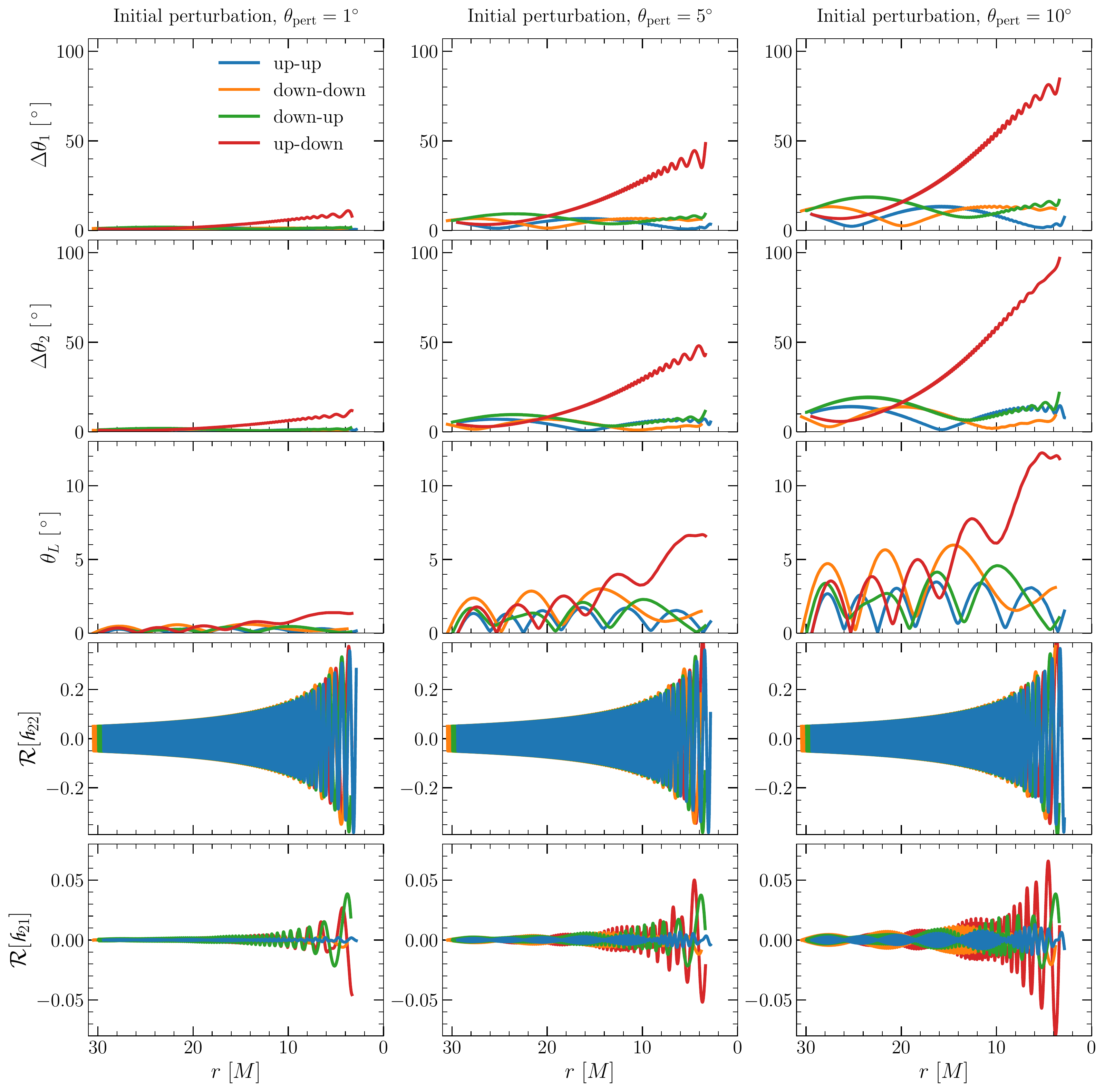}
\caption{
Precessional dynamics and its effect on the gravitational waveform.  Each
subplot shows the four configurations we consider: up-up (blue),
down-down (orange), down-up (green) and up-down (red). Each column corresponds
to a different initial perturbation ($\thetapert$, indicated at the top) of the
spin directions with respect to a perfectly aligned-spin system.  The first
(second) row shows the evolution of the spin perturbation angles
[$\dtheta_{1,2}$; cf. Eq.~(\ref{eq:dtheta})] of the heavier (lighter) BH. The
third row shows the evolution of the orbital plane tilt angle [$\thetaL$; cf.
Eq.~(\ref{eq:thetaL})]. The fourth and fifth rows show the real part of the
(2, 2) and (2, 1) GW modes, respectively. All quantities are shown as a function
of the Newtonian separation $r$ [cf. Eq.~(\ref{eq:separation})] and are
terminated at the separation where the common horizon is found. The instability
development is reflected in a rapid growth of the angles
$\Delta\theta_{1,2}$ and $\thetaL$ for binaries near the up-down configuration.
This in turn impacts the observed GW signal, in particular the subdominant
modes like (2, 1) as the orbital precession causes  power leakage from the
dominant (2, 2) mode.
}
\label{fig:NR_misalignments}
\end{figure*}

\begin{figure*}[thb]
\centering
\includegraphics[width=\linewidth]{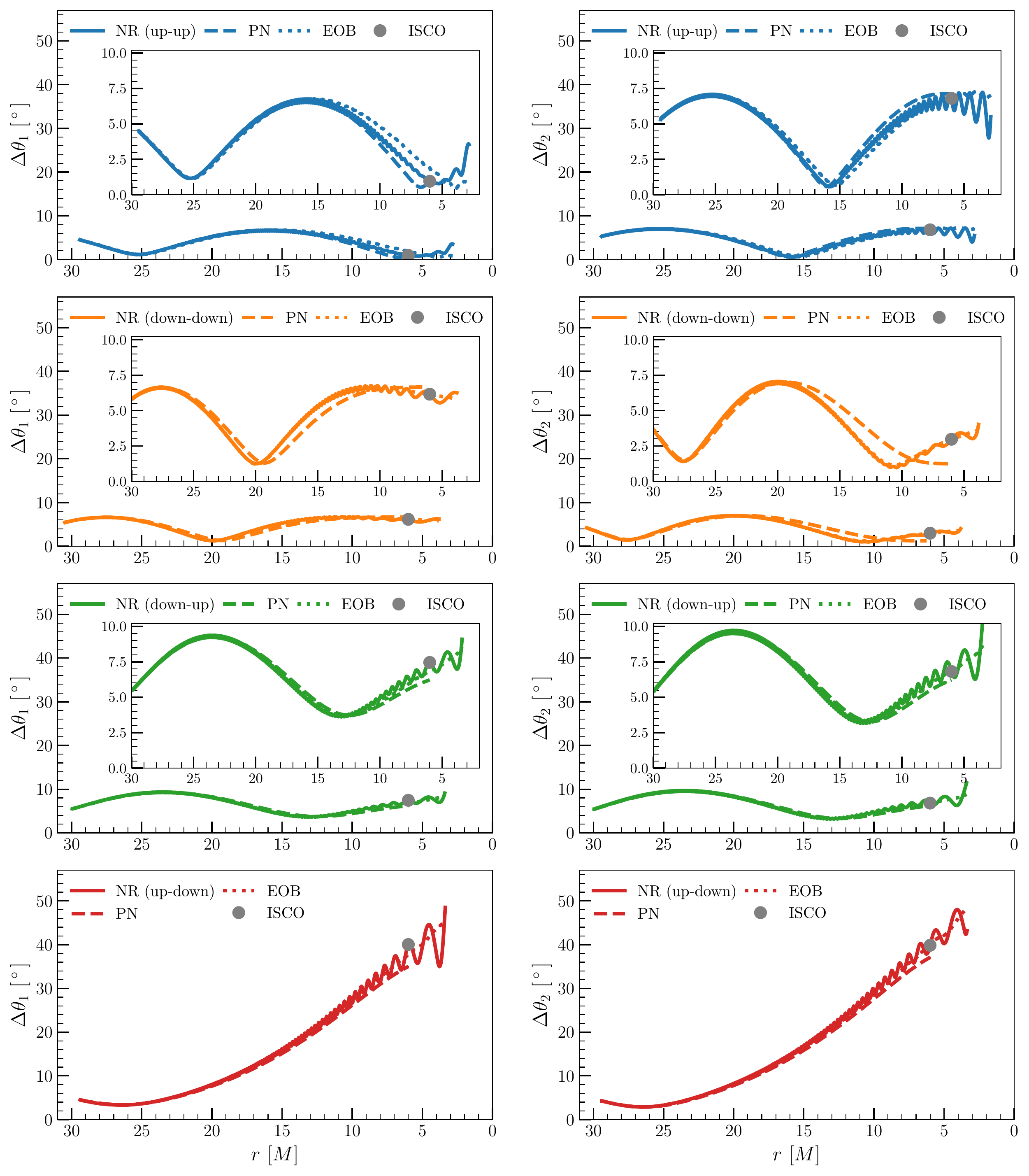}
\caption{
Comparison of the spin dynamics predicted by PN and EOB approximants against
NR. We show the spin perturbation angles $\dtheta_{1,2}$ as a function of the
Newtonian separation $r$ for each of the four configurations: up-up (top-row),
down-down (second-row), down-up (third-row) and up-down (bottom-row), with an
initial perturbation $\thetapert=5\deg$. The left (right) column shows
$\dtheta_1$ ($\dtheta_2$). For the stable configurations, the inset shows an enlarged version of the same panel. The Schwarzschild ISCO radius ($r=6M$) is
indicated by a gray marker. Both approximants are able to track the precession
modulations in $\dtheta_{1,2}$ found in NR, including the unstable growth of
the up-down instability, but fail to match the smaller spin oscillations
occurring on the orbital timescale.
}
\label{fig:NRvsPNandEOB_misalignments}
\end{figure*}

The component BH masses $m_{1}$ and $m_{2}$ and dimensionless spins
$\bchi_{1}(\overline{t})$ and $\bchi_{2}(\overline{t})$ are evaluated on the
apparent horizons~\cite{Boyle:2019kee} of the BHs. Here, $\overline{t}$ is the
simulation time at which the spins are measured in the near
zone~\cite{Boyle:2019kee}. Following previous studies~\cite{Varma:2019csw,
Blackman:2017pcm}, we identify time $t$ with $\overline{t}$.  While this
identification is gauge dependent, the spin directions are already
gauge dependent. However, we note that the spin and orbital angular momentum
vectors in the damped harmonic gauge used by SpEC are in good agreement with the
corresponding PN vectors~\cite{Ossokine:2015vda}.

Following Refs.~\cite{Varma:2019csw, Blackman:2017pcm}, the orbital frequency
$\omegaorb$ is computed as the time derivative of the orbital phase obtained
from the coprecessing-frame waveform at future null infinity. The separation is
then defined as a proxy for the orbital frequency by adopting the Newtonian
expression
\begin{equation}
    r \equiv M \left( M \omegaorb \right)^{-2/3}.
\label{eq:separation}
\end{equation}

We discard the initial $t<t_0=5000M$ of data as this is contaminated by
spurious initial transients caused by imperfect initial data, also known as
junk radiation~\cite{Boyle:2019kee}. This is more stringent than the typical
choice of discarding $~200$--$500M$~\cite{Boyle:2019kee} of data, but we find
that this is necessary to eliminate transient features in $\omegaorb$. In
addition, these long simulations adopt a larger outer
boundary ($R_{\rm{max}}\sim2000M$)~\cite{Szilagyi:2015rwa} compared to typical NR runs ($R_{\rm{max}}\sim600M$)~\cite{Boyle:2019kee}, implying that
junk radiation takes longer to exit the simulation domain. We report the spin
vectors $\bm{\chi}_{1,2}$ and the gravitational waveform $\hlm$ in a frame
where the $z$ direction lies along the orbital angular momentum direction
$\Lhat$ at $t_0$, the $x$ direction is given by the separation vector from the
lighter to the heavier BH at $t_0$, and the $y$ direction completes the
orthonormal triad.  Note that both $\Lhat$ and the separation vector are
estimated using the gravitational waveform $\hlm$ as in
Refs.~\cite{Varma:2019csw, Blackman:2017pcm}.

We compute the tilt angles $\theta_{i}$ for the BH spins ($i=1,2$) as
\begin{gather}
    \cos\theta_{i}(t)=\chihat_{i}(t) \cdot \Lhat(t)\,.
\end{gather}
Their offsets from alignment are given by
\begin{gather}
   \dtheta_{i}(t) \equiv |\theta_{i}(t) -\thetaal|\,,
\label{eq:dtheta}
\end{gather}
where $\thetaal=0\deg \ (180\deg)$ for an unperturbed up (down)
spin component. Finally, we define the instantaneous tilt angle of the orbital
plane, $\thetaL$, as
\begin{gather}
    \cos\thetaL(t) \equiv  \Lhat(t) \cdot \zhat\,.
\label{eq:thetaL}
\end{gather}
Strictly speaking, one has $\dtheta_1=\dtheta_2=\thetapert$ only at the start
of the simulation, but this remains approximately true at $t=t_0$ past  the
junk-radiation stage. Similarly, $\thetaL\simeq0$ at $t_0$. In the following
sections, we use $\dtheta_{1,2}$ and $\thetaL$ to track, respectively, the spin
and orbital plane precession as the binary evolves. The component BH apparent
horizons are tracked until a common apparent horizon is formed at
merger~\cite{Boyle:2019kee}. The variables $\dtheta_{1,2}$ are only available
until this point, which typically corresponds to a separation $r\simeq3M$.

Because our simulations are much longer than the typical simulations performed
with SpEC, their accuracy needs to be investigated separately. We repeat each
simulation with two resolutions, which we refer to as the low- and
high-resolution runs. The grids for these simulations are determined using
adaptive mesh refinement (AMR) as described in Ref.~\cite{Boyle:2019kee} and
references within. The grid resolution varies dynamically during a simulation
to satisfy an AMR tolerance parameter for constraint
violation~\cite{Boyle:2019kee}. For these simulations, the AMR tolerance is
chosen to be a factor of 4 smaller for the high-resolution runs compared
to the low-resolution ones.  Sections~\ref{sec:dynamics} and
\ref{sec:comparisons} present results using the high-resolution runs; the
accuracy of the simulations is then evaluated in Sec.~\ref{sec:nr_resolution}.

\section{Results}
\label{sec:results}

\subsection{Unstable precession dynamics in NR}
\label{sec:dynamics}

Figure~\ref{fig:3d} presents our main result: the up-down precessional
instability, which is a 2PN prediction~\cite{Gerosa:2015tea}, is verified with
full NR.  We show the BH spin evolution for each of the four configurations
with initial perturbation $\thetapert=10\deg$. As the binary inspirals, the BH
spins precess around the $z$ direction, as seen by the purple (orange) curves
which trace the instantaneous values of $\bm{\chi}_1$ ($\bm{\chi}_2$). The
spins of the stable configurations (up-up, down-down and down-up) remain close
to their initial configuration and trace regular precession cones about
alignment. On the other hand, the up-down configuration is very clearly
unstable: the BH spins outspiral dramatically, leading to large misalignments
with respect to $\Lhat$.

While the dominant modulations in the spin directions are the polar
oscillations due to spin precession, much smaller variations, sometimes
referred to as nutations, occur due to the binary orbital motion on the shorter
timescale $\torb\ll\tpre$~\cite{Ossokine:2015vda}. Nutations are negligible
near the beginning of the simulations and they become more pronounced near
merger.

In Fig.~\ref{fig:NR_misalignments} we dissect the unstable dynamics illustrated
in Fig.~\ref{fig:3d} and exhibit their effect on the resulting gravitational
waveforms. For all 12 simulations, we plot various quantities characterizing
the binary dynamics and GW emission as a function of the binary separation $r$
[cf. Eq.~(\ref{eq:separation})].

The top two rows of Fig.~\ref{fig:NR_misalignments} show the spin perturbation
angles $\dtheta_{1,2}$ [cf. Eq.~(\ref{eq:dtheta})] for the component BHs,
indicating the amount of spin precession. First, considering the stable
configurations (up-up, down-down and down-up), as $\thetapert$ is increased
from $1\deg$ (left) to $10\deg$ (right), the amplitudes of the oscillations in
$\dtheta_{1,2}$ increase but
remain close to the initial perturbations. In particular, the maximum
perturbations across all stable configurations are $\max_t \Delta\theta_{1,2}
\simeq 2\deg, 10\deg, 21\deg$ for $\thetapert = 1\deg, 5\deg, 10\deg$,
respectively.

For the up-down binaries, the instability takes a decrease $\sim10M$ in
separation to develop appreciably. Starting from $r\simeq20M$, we observe the
presence of an unstable growth in $\dtheta_{1,2}$. This growth becomes more
rapid with larger initial perturbations $\thetapert$. The maximum perturbations
for the up-down configuration are $\max_t \Delta\theta_{1,2} \simeq 11\deg,
51\deg, 94\deg$ for $\thetapert = 1\deg, 5\deg, 10\deg$, respectively. In
general, the deviations from the initial perturbations increase roughly tenfold
by the end of each simulation (compared to a factor of $\ssim 2$ for the stable
binaries).

These observations are in qualitative agreement with the results presented in
Ref.~\cite{Mould:2020cgc}, which showed with orbit-averaged evolutions at 3.5PN
order that, over a population of up-down binaries, the precessional instability
develops over a typical decrease $\simeq25M$ in the PN orbital separation.  The
study also suggested that the end point of the instability is independent of the
initial perturbation, in contrast to the findings presented here. This is due
to the short nature of NR simulations (though the simulations we performed are
very long by NR standards) and the ``astrophysically unrealistic''
initialization we employed and discussed in Sec.~\ref{sec:nr}. The end point
derivation of Ref.~\cite{Mould:2020cgc} intrinsically relies on binaries
initially close to the up-down configuration at large separations $r>\rudp$
inspiraling through the critical separation before becoming unstable.

In the third row of Fig.~\ref{fig:NR_misalignments}, we show the angle
$\thetaL$ [cf. Eq.~(\ref{eq:thetaL})], which indicates the amount of
orbital-plane precession. The presence of the instability is less apparent in
$\thetaL$ up to at least $r\sim10M$. Even then, the maximum value reached by
$\thetaL$ for the up-down configuration is $\max_t \thetaL \simeq 1\deg, 7\deg,
12\deg$ for $\thetapert = 1\deg, 5\deg, 10\deg$, respectively [though in each
case $\theta_L(t_0)=0\deg$]. The smaller deviation in $\thetaL$ compared to
$\dtheta_{1,2}$ is due to the vastly different magnitudes of the spin and
orbital angular momenta. The binaries simulated here have an initial orbital
frequency $\omegaorb \sim 0.006 \, {\rm rad}/M$ (cf. Table \ref{tab:NRdata})
which corresponds to $L= m_1m_2 (M \omegaorb)^{-1/3}\simeq 1.4M^2$. For $q=0.9$
and $\chi_1=\chi_2=0.8$, the magnitude of the spin angular momenta are given by
$S_1 = m_1^2 \, \chi_{1} \simeq 0.22 M^2$ and $S_2 = m_2^2 \, \chi_{2} \simeq
0.18 M^2$. It is natural to expect that the orbital angular momentum has more
``inertia'' to modulations from the instability and thus
$\theta_L<\Delta\theta_{1,2}$. The angle $\theta_L$  completes $\ssim 3$--$5$
cycles compared to only $\sim 1$ full period for $\theta_{1,2}$. This is in
qualitative agreement with previous PN predictions\footnote{See Fig.~7 in
Ref.~\cite{Zhao:2017tro} for binaries at $r\ssim 10 M$. In their notation, the
ratio between the number of $\Lhat$ cycles and spin-nutation cycles is given by
$\alpha/2\pi$.}~\cite{Zhao:2017tro}.

Notably, orbital precession leads to amplitude modulations of the emitted GW
signal~\cite{Apostolatos:1994pre}.  In the bottom two rows of
Fig.~\ref{fig:NR_misalignments}, we investigate the influence of the up-down
precessional instability on the gravitational waveforms; we focus in particular
on the dominant $(2,2)$ mode and a subdominant mode $(2,1)$ of the GW strain
decomposition given in Eq.~(\ref{eq:spherical_harm}).

The $(2,2)$ mode is largely unaffected by the sensitive details of precession --the simulations of all four configurations and three initial perturbations
present qualitatively similar results in this mode.  The dominant morphological
features of the $(2,2)$ mode waveform are the GW cycles at approximately twice
the orbital frequency and the typical ``chirp'' as the binary merges.  However,
we find that even the modest growth of $\thetaL$ mentioned above leaves a
notable imprint on the subdominant modes like the $(2,1)$ mode.  This is
apparent by comparing the third ($\thetaL$) and fifth ($\h_{21}$) rows of
Fig.~\ref{fig:NR_misalignments}. Starting from $r\ssim10M$, as $\thetaL$
experiences a growth for the up-down binary (indicating precession of the
orbital plane), so does the amplitude of the $(2,1)$ mode. This arises because
orbital precession induces a transfer of power from the $(2,2)$ mode to the
subdominant modes (see, e.g., \cite{Varma:2019csw}).  This transfer of power to
subdominant modes is a feature of generic precessing binaries, not just the
peculiar up-down configuration.  The growth in $\theta_L$ becomes more
pronounced as we increase the initial perturbation $\thetapert$ from $1\deg$ to
$10\deg$.  For $\thetapert=10\deg$, the amplitude of the $(2,1)$ mode near
merger is about twice as large for the up-down configuration compared to the
three stable binaries. We note that, for astrophysically realistic binaries, as
they undergo the instability at much larger separations, we expect an even
larger imprint on the waveform.

\begin{figure*}[thb]
\centering
\includegraphics[width=\linewidth]{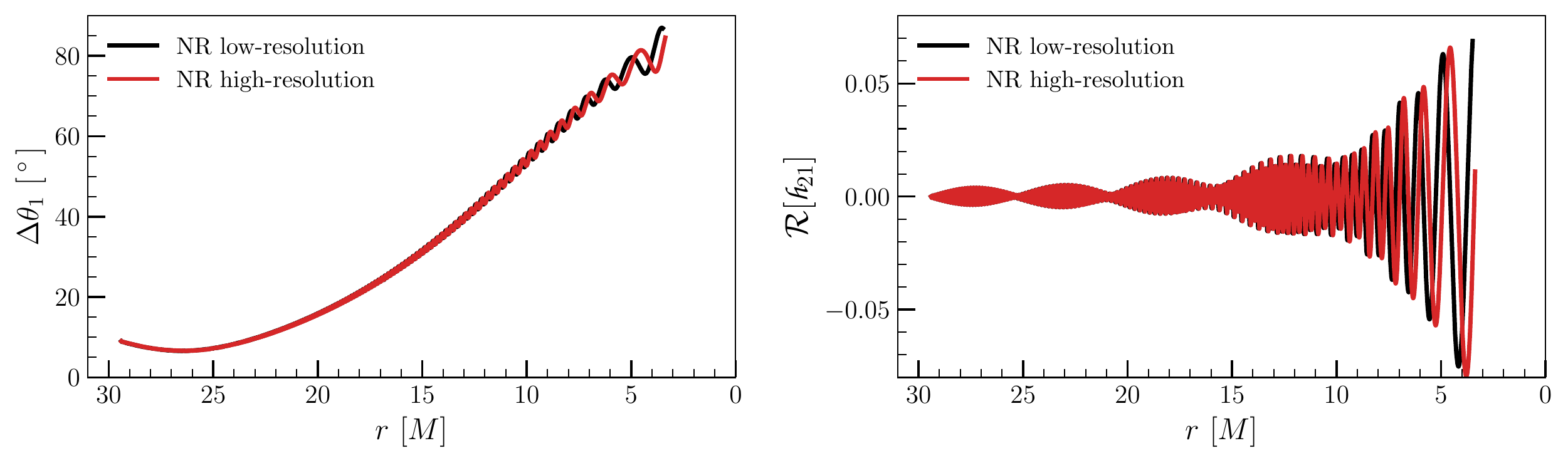}
\caption{
Comparison of the spin dynamics and gravitational waveform between simulations
with two different NR resolutions for the up-down system with an initial
perturbation $\thetapert=10\deg$. We show the spin perturbation angle
$\dtheta_{1}$ and the real part of the (2, 1) mode, as a function of the
Newtonian separation $r$.
}
\label{fig:NR_resolution}
\end{figure*}

\subsection{Comparison with approximate evolutions}
\label{sec:comparisons}

We now test how well approximate PN and EOB methods capture the full NR
dynamics for stable and unstable aligned-spin binaries. The PN dynamics is
evaluated using the \texttt{SpinTaylorT1}~\cite{Buonanno:2009zt,
Isoyama:2020lls} approximant at 3.5PN as implemented in the LIGO Algorithm
Library (LAL) Suite~\cite{lalsuite}.  We find that \texttt{SpinTaylorT1} is
marginally more accurate than the other available approximants,
\texttt{SpinTaylorT4}~\cite{Buonanno:2002fy} and
\texttt{SpinTaylorT5}~\cite{Ajith:2011ec}. The EOB dynamics is evaluated using
the \texttt{SEOBNRv4PHM} model~\cite{Ossokine:2020kjp}. The evolutions are
initialized using spins and orbital frequency extracted from the NR simulations
at $t_0$. PN evolutions are terminated near the Schwarzschild innermost stable
circular orbit (ISCO), which is located at $r=6M$. Although the EOB model
includes merger and ringdown, we evaluate it only until the NR frequency at
which a common horizon forms.

Figure~\ref{fig:NRvsPNandEOB_misalignments} compares the spin perturbation
angles $\dtheta_{1,2}$ of the PN, EOB, and NR evolutions for each of the four
aligned-spin configurations with $\thetapert=5\deg$. Notably, we find that both
the PN and EOB evolutions reproduce the unstable growth of up-down binaries
(bottom row of Fig.~\ref{fig:NRvsPNandEOB_misalignments}). As expected, the
faithfulness of approximate evolutions decreases at smaller separations, where
the gravitational interaction is strongest and highly nonlinear. The
Schwarzschild ISCO provides a simple proxy to characterize this transition
(gray markers in Fig.~\ref{fig:NRvsPNandEOB_misalignments}).

Overall, we report discrepancies in $\dtheta_{1,2}$ between NR and PN/EOB  of
$\simeq5\deg$ for up-down  and $\simeq1\deg$ for the three stable
configurations (up-up, down-down, and down-up).  EOB evolutions appear to
reproduce the full NR dynamics more accurately. This is perhaps expected since
the EOB framework receives NR information, although it is important to point
out that its calibration does not make use of simulations with precessing
spins. More specifically, in the up-down case,
while the EOB curve follows the orbit-averaged value of the NR evolution,
the PN curve tracks the minimum of each orbital cycle. Neither method matches
the orbital timescale modulations present in the NR simulations (cf.
Refs.~\cite{Ossokine:2015vda, Owen:2017yaj} for related work).

\subsection{Impact of NR resolution error}
\label{sec:nr_resolution}

The results presented so far were based on our high-resolution NR simulations.
We now investigate the impact of the numerical resolution error on our results
by comparing the output of our low- and high-resolution simulations (cf.
Sec.~\ref{sec:nr}). Figure~\ref{fig:NR_resolution} compares two of the main
quantities of interest in this work: the spin perturbation angle $\dtheta_{1}$
($\dtheta_2$ is qualitatively similar) and the (2, 1) GW mode, for the up-down
case with $\thetapert=10\deg$.  The agreement between the two resolutions
slowly degrades as the binary approaches merger, accumulating a dephasing of
$\sim0.9$ rad over a phase evolution of $\sim290$ rad in the (2, 1) mode. This
suggests that even higher resolution simulations might be necessary to fully
capture the fine details of the instability. However, all the key features such
as the growth of the instability in $\dtheta_{1,2}$, magnitude of the orbital
timescale oscillations in $\dtheta_{1,2}$, and the growth of the (2,1) mode,
are well captured and qualitatively similar between the two sets of runs.

\section{Conclusions}
\label{sec:conclusions}

We presented the first NR simulations of aligned-spin binary BHs undergoing a
precessional instability, verifying that previous PN
predictions~\cite{Gerosa:2015hba} hold in the strong-field regime of GR. The
instability occurs for binary BHs in the up-down configuration, where the spin
of the heavier (lighter) BH is aligned (antialigned) with the orbital angular
momentum. Initialized with a small spin misalignment, as the binary evolves,
the instability causes the spins to tilt dramatically from their initial
configuration, achieving misalignments up to $\ssim90\deg$ at merger. In order
to observe this precessional effect, the simulations we perform are necessarily
long. Each consists of $\ssim100$ orbits and lasts for a time
$\sim65\times10^3M$, putting them among the longest simulations in the SXS
catalog~\cite{SXSWebsite, Boyle:2019kee}, which previously had just 12
simulations with $>90$ orbits.

We perform three simulations of up-down binaries, each with mass ratio $q=0.9$,
dimensionless spins $\chi_1 = \chi_2 =0.8$, and increasing initial misalignment
$\thetapert = 1\deg, 5\deg, 10\deg$. All three exhibit the unstable precession
behavior, while the rate of growth of the instability increases with the
initial misalignment. We show that the instability leaves a strong imprint on
the subdominant modes of the GW signal in the up-down binaries, which can
potentially be used to distinguish them from other sources. We repeat these
simulations for the other three aligned-spin configurations (up-up, down-down,
and down-up) with the same parameters and show that they all remain stable,
undergoing only small-angle precession oscillations.

We compare the results of our NR simulations against both
PN~\cite{Buonanno:2009zt} and EOB~\cite{Ossokine:2020kjp} evolutions. We find
that both frameworks capture the occurrence and growth of the up-down
instability. While the EOB dynamics is more accurate in predicting the
precession-timescale oscillations, both methods fail to match the
orbital-timescale modulations seen in NR. Current NR surrogate
models~\cite{Varma:2019csw, Blackman:2017pcm} are unable to reproduce the
up-down instability because they are trained on NR simulations that last only
$\sim20$ orbits, during which the instability does not have time to develop.
It would be interesting to see if current surrogate techniques can indeed
capture this instability if applied to hybridized EOB-NR waveforms (cf.
Ref.~\cite{Varma:2018mmi} for work in this direction).

Through the up-down instability, binary BHs whose astrophysical formation leads
to spins that are initially (anti) aligned with the orbital angular momentum
can become misaligned and strongly precessing near merger. Whether current
LIGO/Virgo parameter-estimation techniques can confidently identify unstable
up-down binaries as such remains an open point of investigation.

\begin{acknowledgments}
We thank Daria Gangardt, Serguei Ossokine, Ulrich Sperhake, and Richard O'Shaughnessy for useful discussions.
V.V.\ is supported by a Klarman Fellowship at Cornell and National Science Foundation (NSF) Grants No. PHY-170212 and No. PHY-1708213 at
Caltech.
D.G. and M.M. are supported by European Union's H2020  ERC Starting Grant No.
945155-GWmining and  Royal Society Grant No. RGS-R2-202004. D.G. is supported
by Leverhulme Trust Grant No. RPG-2019-350.
V.V., M.A.S., and L.E.K. are supported by the Sherman Fairchild Foundation.
M.A.S. is supported by NSF Grants No.
PHY-2011961, No. PHY-2011968, and No. OAC-1931266 at Caltech.
L.E.K. is supported NSF Grants No.
PHY-1912081 and No. OAC-1931280 at Cornell.
Simulations for this work were performed on the Frontera
cluster~\cite{Frontera}, which is supported by the Texas Advanced Computing
Center (TACC) at the University of Texas at Austin.
Additional computational work was performed on the Wheeler cluster at Caltech,
which is supported by the Sherman Fairchild Foundation and Caltech, the
University of Birmingham BlueBEAR cluster, the Athena cluster at HPC Midlands+
funded by Engineering and Physical Sciences Research Council Grant No.  EP/P020232/1, and the Maryland Advanced Research
Computing Center (MARCC).

\end{acknowledgments}

\bibliography{references}

\begin{thebibliography}{50}%
\makeatletter
\providecommand \@ifxundefined [1]{%
 \@ifx{#1\undefined}
}%
\providecommand \@ifnum [1]{%
 \ifnum #1\expandafter \@firstoftwo
 \else \expandafter \@secondoftwo
 \fi
}%
\providecommand \@ifx [1]{%
 \ifx #1\expandafter \@firstoftwo
 \else \expandafter \@secondoftwo
 \fi
}%
\providecommand \natexlab [1]{#1}%
\providecommand \enquote  [1]{``#1''}%
\providecommand \bibnamefont  [1]{#1}%
\providecommand \bibfnamefont [1]{#1}%
\providecommand \citenamefont [1]{#1}%
\providecommand \href@noop [0]{\@secondoftwo}%
\providecommand \href [0]{\begingroup \@sanitize@url \@href}%
\providecommand \@href[1]{\@@startlink{#1}\@@href}%
\providecommand \@@href[1]{\endgroup#1\@@endlink}%
\providecommand \@sanitize@url [0]{\catcode `\\12\catcode `\$12\catcode
  `\&12\catcode `\#12\catcode `\^12\catcode `\_12\catcode `\%12\relax}%
\providecommand \@@startlink[1]{}%
\providecommand \@@endlink[0]{}%
\providecommand \url  [0]{\begingroup\@sanitize@url \@url }%
\providecommand \@url [1]{\endgroup\@href {#1}{\urlprefix }}%
\providecommand \urlprefix  [0]{URL }%
\providecommand \Eprint [0]{\href }%
\providecommand \doibase [0]{http://dx.doi.org/}%
\providecommand \selectlanguage [0]{\@gobble}%
\providecommand \bibinfo  [0]{\@secondoftwo}%
\providecommand \bibfield  [0]{\@secondoftwo}%
\providecommand \translation [1]{[#1]}%
\providecommand \BibitemOpen [0]{}%
\providecommand \bibitemStop [0]{}%
\providecommand \bibitemNoStop [0]{.\EOS\space}%
\providecommand \EOS [0]{\spacefactor3000\relax}%
\providecommand \BibitemShut  [1]{\csname bibitem#1\endcsname}%
\let\auto@bib@innerbib\@empty
%</preamble>
\bibitem [{\citenamefont {Abbott}\ \emph {et~al.}(2019)\citenamefont {Abbott}
  \emph {et~al.}}]{LIGOScientific:2018mvr}%
  \BibitemOpen
  \bibfield  {author} {\bibinfo {author} {\bibfnamefont {B.~P.}\ \bibnamefont
  {Abbott}} \emph {et~al.} (\bibinfo {collaboration} {LIGO Scientific,
  Virgo}),\ }\bibfield  {title} {\enquote {\bibinfo {title} {{GWTC-1: A
  Gravitational-Wave Transient Catalog of Compact Binary Mergers Observed by
  LIGO and Virgo during the First and Second Observing Runs}},}\ }\href
  {\doibase 10.1103/PhysRevX.9.031040} {\bibfield  {journal} {\bibinfo
  {journal} {Phys. Rev.}\ }\textbf {\bibinfo {volume} {X9}},\ \bibinfo {pages}
  {031040} (\bibinfo {year} {2019})},\ \Eprint
  {http://arxiv.org/abs/1811.12907} {arXiv:1811.12907 [astro-ph.HE]}
  \BibitemShut {NoStop}%
%%CITATION = ARXIV:1811.12907;%%
\bibitem [{\citenamefont {Abbott}\ \emph
  {et~al.}(2020{\natexlab{a}})\citenamefont {Abbott} \emph
  {et~al.}}]{Abbott:2020niy}%
  \BibitemOpen
  \bibfield  {author} {\bibinfo {author} {\bibfnamefont {R.}~\bibnamefont
  {Abbott}} \emph {et~al.} (\bibinfo {collaboration} {LIGO Scientific,
  Virgo}),\ }\bibfield  {title} {\enquote {\bibinfo {title} {{GWTC-2: Compact
  Binary Coalescences Observed by LIGO and Virgo During the First Half of the
  Third Observing Run}},}\ }\href@noop {} {\  (\bibinfo {year}
  {2020}{\natexlab{a}})},\ \Eprint {http://arxiv.org/abs/2010.14527}
  {arXiv:2010.14527 [gr-qc]} \BibitemShut {NoStop}%
\bibitem [{\citenamefont {Abbott}\ \emph {et~al.}(2018)\citenamefont {Abbott}
  \emph {et~al.}}]{LIGOScientific:2018jsj}%
  \BibitemOpen
  \bibfield  {author} {\bibinfo {author} {\bibfnamefont {B.~P.}\ \bibnamefont
  {Abbott}} \emph {et~al.} (\bibinfo {collaboration} {LIGO Scientific,
  Virgo}),\ }\bibfield  {title} {\enquote {\bibinfo {title} {{Binary Black Hole
  Population Properties Inferred from the First and Second Observing Runs of
  Advanced LIGO and Advanced Virgo}},}\ }\href@noop {} {\  (\bibinfo {year}
  {2018})},\ \Eprint {http://arxiv.org/abs/1811.12940} {arXiv:1811.12940
  [astro-ph.HE]} \BibitemShut {NoStop}%
%%CITATION = ARXIV:1811.12940;%%
\bibitem [{\citenamefont {Abbott}\ \emph
  {et~al.}(2020{\natexlab{b}})\citenamefont {Abbott} \emph
  {et~al.}}]{Abbott:2020gyp}%
  \BibitemOpen
  \bibfield  {author} {\bibinfo {author} {\bibfnamefont {R.}~\bibnamefont
  {Abbott}} \emph {et~al.} (\bibinfo {collaboration} {LIGO Scientific,
  Virgo}),\ }\bibfield  {title} {\enquote {\bibinfo {title} {{Population
  Properties of Compact Objects from the Second LIGO-Virgo Gravitational-Wave
  Transient Catalog}},}\ }\href@noop {} {\  (\bibinfo {year}
  {2020}{\natexlab{b}})},\ \Eprint {http://arxiv.org/abs/2010.14533}
  {arXiv:2010.14533 [astro-ph.HE]} \BibitemShut {NoStop}%
\bibitem [{\citenamefont {Roulet}\ \emph {et~al.}(2020)\citenamefont {Roulet},
  \citenamefont {Venumadhav}, \citenamefont {Zackay}, \citenamefont {Dai},\
  and\ \citenamefont {Zaldarriaga}}]{Roulet:2020wyq}%
  \BibitemOpen
  \bibfield  {author} {\bibinfo {author} {\bibfnamefont {Javier}\ \bibnamefont
  {Roulet}}, \bibinfo {author} {\bibfnamefont {Tejaswi}\ \bibnamefont
  {Venumadhav}}, \bibinfo {author} {\bibfnamefont {Barak}\ \bibnamefont
  {Zackay}}, \bibinfo {author} {\bibfnamefont {Liang}\ \bibnamefont {Dai}}, \
  and\ \bibinfo {author} {\bibfnamefont {Matias}\ \bibnamefont {Zaldarriaga}},\
  }\bibfield  {title} {\enquote {\bibinfo {title} {{Binary Black Hole Mergers
  from LIGO/Virgo O1 and O2: Population Inference Combining Confident and
  Marginal Events}},}\ }\href@noop {} {\  (\bibinfo {year} {2020})},\ \Eprint
  {http://arxiv.org/abs/2008.07014} {arXiv:2008.07014 [astro-ph.HE]}
  \BibitemShut {NoStop}%
%%CITATION = ARXIV:2008.07014;%%
\bibitem [{\citenamefont {Apostolatos}\ \emph {et~al.}(1994)\citenamefont
  {Apostolatos}, \citenamefont {Cutler}, \citenamefont {Sussman},\ and\
  \citenamefont {Thorne}}]{Apostolatos:1994pre}%
  \BibitemOpen
  \bibfield  {author} {\bibinfo {author} {\bibfnamefont {Theocharis~A.}\
  \bibnamefont {Apostolatos}}, \bibinfo {author} {\bibfnamefont {Curt}\
  \bibnamefont {Cutler}}, \bibinfo {author} {\bibfnamefont {Gerald~J.}\
  \bibnamefont {Sussman}}, \ and\ \bibinfo {author} {\bibfnamefont {Kip~S.}\
  \bibnamefont {Thorne}},\ }\bibfield  {title} {\enquote {\bibinfo {title}
  {Spin-induced orbital precession and its modulation of the gravitational
  waveforms from merging binaries},}\ }\href {\doibase
  10.1103/PhysRevD.49.6274} {\bibfield  {journal} {\bibinfo  {journal} {Phys.
  Rev. D}\ }\textbf {\bibinfo {volume} {49}},\ \bibinfo {pages} {6274--6297}
  (\bibinfo {year} {1994})}\BibitemShut {NoStop}%
\bibitem [{\citenamefont {Kidder}(1995)}]{Kidder:1995zr}%
  \BibitemOpen
  \bibfield  {author} {\bibinfo {author} {\bibfnamefont {Lawrence~E.}\
  \bibnamefont {Kidder}},\ }\bibfield  {title} {\enquote {\bibinfo {title}
  {{Coalescing binary systems of compact objects to postNewtonian 5/2 order. 5.
  Spin effects}},}\ }\href {\doibase 10.1103/PhysRevD.52.821} {\bibfield
  {journal} {\bibinfo  {journal} {Phys. Rev. D}\ }\textbf {\bibinfo {volume}
  {52}},\ \bibinfo {pages} {821--847} (\bibinfo {year} {1995})},\ \Eprint
  {http://arxiv.org/abs/gr-qc/9506022} {arXiv:gr-qc/9506022} \BibitemShut
  {NoStop}%
\bibitem [{\citenamefont {Gerosa}\ \emph {et~al.}(2013)\citenamefont {Gerosa},
  \citenamefont {Kesden}, \citenamefont {Berti}, \citenamefont
  {O'Shaughnessy},\ and\ \citenamefont {Sperhake}}]{Gerosa:2013laa}%
  \BibitemOpen
  \bibfield  {author} {\bibinfo {author} {\bibfnamefont {Davide}\ \bibnamefont
  {Gerosa}}, \bibinfo {author} {\bibfnamefont {Michael}\ \bibnamefont
  {Kesden}}, \bibinfo {author} {\bibfnamefont {Emanuele}\ \bibnamefont
  {Berti}}, \bibinfo {author} {\bibfnamefont {Richard}\ \bibnamefont
  {O'Shaughnessy}}, \ and\ \bibinfo {author} {\bibfnamefont {Ulrich}\
  \bibnamefont {Sperhake}},\ }\bibfield  {title} {\enquote {\bibinfo {title}
  {{Resonant-plane locking and spin alignment in stellar-mass black-hole
  binaries: a diagnostic of compact-binary formation}},}\ }\href {\doibase
  10.1103/PhysRevD.87.104028} {\bibfield  {journal} {\bibinfo  {journal} {Phys.
  Rev.}\ }\textbf {\bibinfo {volume} {D87}},\ \bibinfo {pages} {104028}
  (\bibinfo {year} {2013})},\ \Eprint {http://arxiv.org/abs/1302.4442}
  {arXiv:1302.4442 [gr-qc]} \BibitemShut {NoStop}%
%%CITATION = ARXIV:1302.4442;%%
\bibitem [{\citenamefont {Rodriguez}\ \emph {et~al.}(2016)\citenamefont
  {Rodriguez}, \citenamefont {Zevin}, \citenamefont {Pankow}, \citenamefont
  {Kalogera},\ and\ \citenamefont {Rasio}}]{Rodriguez:2016vmx}%
  \BibitemOpen
  \bibfield  {author} {\bibinfo {author} {\bibfnamefont {Carl~L.}\ \bibnamefont
  {Rodriguez}}, \bibinfo {author} {\bibfnamefont {Michael}\ \bibnamefont
  {Zevin}}, \bibinfo {author} {\bibfnamefont {Chris}\ \bibnamefont {Pankow}},
  \bibinfo {author} {\bibfnamefont {Vasilliki}\ \bibnamefont {Kalogera}}, \
  and\ \bibinfo {author} {\bibfnamefont {Frederic~A.}\ \bibnamefont {Rasio}},\
  }\bibfield  {title} {\enquote {\bibinfo {title} {{Illuminating Black Hole
  Binary Formation Channels with Spins in Advanced LIGO}},}\ }\href {\doibase
  10.3847/2041-8205/832/1/L2} {\bibfield  {journal} {\bibinfo  {journal}
  {Astrophys. J. Lett.}\ }\textbf {\bibinfo {volume} {832}},\ \bibinfo {pages}
  {L2} (\bibinfo {year} {2016})},\ \Eprint {http://arxiv.org/abs/1609.05916}
  {arXiv:1609.05916 [astro-ph.HE]} \BibitemShut {NoStop}%
\bibitem [{\citenamefont {Vitale}\ \emph {et~al.}(2017)\citenamefont {Vitale},
  \citenamefont {Lynch}, \citenamefont {Sturani},\ and\ \citenamefont
  {Graff}}]{Vitale:2015tea}%
  \BibitemOpen
  \bibfield  {author} {\bibinfo {author} {\bibfnamefont {Salvatore}\
  \bibnamefont {Vitale}}, \bibinfo {author} {\bibfnamefont {Ryan}\ \bibnamefont
  {Lynch}}, \bibinfo {author} {\bibfnamefont {Riccardo}\ \bibnamefont
  {Sturani}}, \ and\ \bibinfo {author} {\bibfnamefont {Philip}\ \bibnamefont
  {Graff}},\ }\bibfield  {title} {\enquote {\bibinfo {title} {{Use of
  gravitational waves to probe the formation channels of compact binaries}},}\
  }\href {\doibase 10.1088/1361-6382/aa552e} {\bibfield  {journal} {\bibinfo
  {journal} {Class. Quant. Grav.}\ }\textbf {\bibinfo {volume} {34}},\ \bibinfo
  {pages} {03LT01} (\bibinfo {year} {2017})},\ \Eprint
  {http://arxiv.org/abs/1503.04307} {arXiv:1503.04307 [gr-qc]} \BibitemShut
  {NoStop}%
%%CITATION = ARXIV:1503.04307;%%
\bibitem [{\citenamefont {Talbot}\ and\ \citenamefont
  {Thrane}(2017)}]{Talbot:2017yur}%
  \BibitemOpen
  \bibfield  {author} {\bibinfo {author} {\bibfnamefont {Colm}\ \bibnamefont
  {Talbot}}\ and\ \bibinfo {author} {\bibfnamefont {Eric}\ \bibnamefont
  {Thrane}},\ }\bibfield  {title} {\enquote {\bibinfo {title} {{Determining the
  population properties of spinning black holes}},}\ }\href {\doibase
  10.1103/PhysRevD.96.023012} {\bibfield  {journal} {\bibinfo  {journal} {Phys.
  Rev. D}\ }\textbf {\bibinfo {volume} {96}},\ \bibinfo {pages} {023012}
  (\bibinfo {year} {2017})},\ \Eprint {http://arxiv.org/abs/1704.08370}
  {arXiv:1704.08370 [astro-ph.HE]} \BibitemShut {NoStop}%
\bibitem [{\citenamefont {Farr}\ \emph {et~al.}(2017)\citenamefont {Farr},
  \citenamefont {Stevenson}, \citenamefont {Coleman~Miller}, \citenamefont
  {Mandel}, \citenamefont {Farr},\ and\ \citenamefont
  {Vecchio}}]{Farr:2017uvj}%
  \BibitemOpen
  \bibfield  {author} {\bibinfo {author} {\bibfnamefont {Will~M.}\ \bibnamefont
  {Farr}}, \bibinfo {author} {\bibfnamefont {Simon}\ \bibnamefont {Stevenson}},
  \bibinfo {author} {\bibfnamefont {M.}~\bibnamefont {Coleman~Miller}},
  \bibinfo {author} {\bibfnamefont {Ilya}\ \bibnamefont {Mandel}}, \bibinfo
  {author} {\bibfnamefont {Ben}\ \bibnamefont {Farr}}, \ and\ \bibinfo {author}
  {\bibfnamefont {Alberto}\ \bibnamefont {Vecchio}},\ }\bibfield  {title}
  {\enquote {\bibinfo {title} {{Distinguishing Spin-Aligned and Isotropic Black
  Hole Populations With Gravitational Waves}},}\ }\href {\doibase
  10.1038/nature23453} {\bibfield  {journal} {\bibinfo  {journal} {Nature}\
  }\textbf {\bibinfo {volume} {548}},\ \bibinfo {pages} {426} (\bibinfo {year}
  {2017})},\ \Eprint {http://arxiv.org/abs/1706.01385} {arXiv:1706.01385
  [astro-ph.HE]} \BibitemShut {NoStop}%
\bibitem [{\citenamefont {Stevenson}\ \emph {et~al.}(2017)\citenamefont
  {Stevenson}, \citenamefont {Berry},\ and\ \citenamefont
  {Mandel}}]{Stevenson:2017dlk}%
  \BibitemOpen
  \bibfield  {author} {\bibinfo {author} {\bibfnamefont {Simon}\ \bibnamefont
  {Stevenson}}, \bibinfo {author} {\bibfnamefont {Christopher~P.L.}\
  \bibnamefont {Berry}}, \ and\ \bibinfo {author} {\bibfnamefont {Ilya}\
  \bibnamefont {Mandel}},\ }\bibfield  {title} {\enquote {\bibinfo {title}
  {{Hierarchical analysis of gravitational-wave measurements of binary black
  hole spin\textendash{}orbit misalignments}},}\ }\href {\doibase
  10.1093/mnras/stx1764} {\bibfield  {journal} {\bibinfo  {journal} {Mon. Not.
  Roy. Astron. Soc.}\ }\textbf {\bibinfo {volume} {471}},\ \bibinfo {pages}
  {2801--2811} (\bibinfo {year} {2017})},\ \Eprint
  {http://arxiv.org/abs/1703.06873} {arXiv:1703.06873 [astro-ph.HE]}
  \BibitemShut {NoStop}%
\bibitem [{\citenamefont {Gerosa}\ \emph {et~al.}(2018)\citenamefont {Gerosa},
  \citenamefont {Berti}, \citenamefont {O'Shaughnessy}, \citenamefont
  {Belczynski}, \citenamefont {Kesden}, \citenamefont {Wysocki},\ and\
  \citenamefont {Gladysz}}]{Gerosa:2018wbw}%
  \BibitemOpen
  \bibfield  {author} {\bibinfo {author} {\bibfnamefont {Davide}\ \bibnamefont
  {Gerosa}}, \bibinfo {author} {\bibfnamefont {Emanuele}\ \bibnamefont
  {Berti}}, \bibinfo {author} {\bibfnamefont {Richard}\ \bibnamefont
  {O'Shaughnessy}}, \bibinfo {author} {\bibfnamefont {Krzysztof}\ \bibnamefont
  {Belczynski}}, \bibinfo {author} {\bibfnamefont {Michael}\ \bibnamefont
  {Kesden}}, \bibinfo {author} {\bibfnamefont {Daniel}\ \bibnamefont
  {Wysocki}}, \ and\ \bibinfo {author} {\bibfnamefont {Wojciech}\ \bibnamefont
  {Gladysz}},\ }\bibfield  {title} {\enquote {\bibinfo {title} {{Spin
  orientations of merging black holes formed from the evolution of stellar
  binaries}},}\ }\href {\doibase 10.1103/PhysRevD.98.084036} {\bibfield
  {journal} {\bibinfo  {journal} {Phys. Rev.}\ }\textbf {\bibinfo {volume}
  {D98}},\ \bibinfo {pages} {084036} (\bibinfo {year} {2018})},\ \Eprint
  {http://arxiv.org/abs/1808.02491} {arXiv:1808.02491 [astro-ph.HE]}
  \BibitemShut {NoStop}%
%%CITATION = ARXIV:1808.02491;%%
\bibitem [{\citenamefont {Belczynski}\ \emph {et~al.}(2020)\citenamefont
  {Belczynski} \emph {et~al.}}]{Belczynski:2017gds}%
  \BibitemOpen
  \bibfield  {author} {\bibinfo {author} {\bibfnamefont {K.}~\bibnamefont
  {Belczynski}} \emph {et~al.},\ }\bibfield  {title} {\enquote {\bibinfo
  {title} {{Evolutionary roads leading to low effective spins, high black hole
  masses, and O1/O2 rates for LIGO/Virgo binary black holes}},}\ }\href
  {\doibase 10.1051/0004-6361/201936528} {\bibfield  {journal} {\bibinfo
  {journal} {Astron. Astrophys.}\ }\textbf {\bibinfo {volume} {636}},\ \bibinfo
  {pages} {A104} (\bibinfo {year} {2020})},\ \Eprint
  {http://arxiv.org/abs/1706.07053} {arXiv:1706.07053 [astro-ph.HE]}
  \BibitemShut {NoStop}%
%%CITATION = ARXIV:1706.07053;%%
\bibitem [{\citenamefont {Kalogera}(2000)}]{Kalogera:1999tq}%
  \BibitemOpen
  \bibfield  {author} {\bibinfo {author} {\bibfnamefont {Vassiliki}\
  \bibnamefont {Kalogera}},\ }\bibfield  {title} {\enquote {\bibinfo {title}
  {{Spin orbit misalignment in close binaries with two compact objects}},}\
  }\href {\doibase 10.1086/309400} {\bibfield  {journal} {\bibinfo  {journal}
  {Astrophys. J.}\ }\textbf {\bibinfo {volume} {541}},\ \bibinfo {pages}
  {319--328} (\bibinfo {year} {2000})},\ \Eprint
  {http://arxiv.org/abs/astro-ph/9911417} {arXiv:astro-ph/9911417} \BibitemShut
  {NoStop}%
\bibitem [{\citenamefont {Yang}\ \emph
  {et~al.}(2019{\natexlab{a}})\citenamefont {Yang}, \citenamefont {Bartos},
  \citenamefont {Haiman}, \citenamefont {Kocsis}, \citenamefont {Marka},
  \citenamefont {Stone},\ and\ \citenamefont {Marka}}]{Yang:2019okq}%
  \BibitemOpen
  \bibfield  {author} {\bibinfo {author} {\bibfnamefont {Y.}~\bibnamefont
  {Yang}}, \bibinfo {author} {\bibfnamefont {I.}~\bibnamefont {Bartos}},
  \bibinfo {author} {\bibfnamefont {Z.}~\bibnamefont {Haiman}}, \bibinfo
  {author} {\bibfnamefont {B.}~\bibnamefont {Kocsis}}, \bibinfo {author}
  {\bibfnamefont {Z.}~\bibnamefont {Marka}}, \bibinfo {author} {\bibfnamefont
  {N.~C.}\ \bibnamefont {Stone}}, \ and\ \bibinfo {author} {\bibfnamefont
  {S.}~\bibnamefont {Marka}},\ }\bibfield  {title} {\enquote {\bibinfo {title}
  {{AGN Disks Harden the Mass Distribution of Stellar-mass Binary Black Hole
  Mergers}},}\ }\href {\doibase 10.3847/1538-4357/ab16e3} {\bibfield  {journal}
  {\bibinfo  {journal} {Astrophys. J.}\ }\textbf {\bibinfo {volume} {876}},\
  \bibinfo {pages} {122} (\bibinfo {year} {2019}{\natexlab{a}})},\ \Eprint
  {http://arxiv.org/abs/1903.01405} {arXiv:1903.01405 [astro-ph.HE]}
  \BibitemShut {NoStop}%
%%CITATION = ARXIV:1903.01405;%%
\bibitem [{\citenamefont {McKernan}\ \emph {et~al.}(2020)\citenamefont
  {McKernan}, \citenamefont {Ford}, \citenamefont {O'Shaughnessy},\ and\
  \citenamefont {Wysocki}}]{McKernan:2019beu}%
  \BibitemOpen
  \bibfield  {author} {\bibinfo {author} {\bibfnamefont {B.}~\bibnamefont
  {McKernan}}, \bibinfo {author} {\bibfnamefont {K.~E.~S.}\ \bibnamefont
  {Ford}}, \bibinfo {author} {\bibfnamefont {R.}~\bibnamefont {O'Shaughnessy}},
  \ and\ \bibinfo {author} {\bibfnamefont {D.}~\bibnamefont {Wysocki}},\
  }\bibfield  {title} {\enquote {\bibinfo {title} {{Monte Carlo simulations of
  black hole mergers in AGN discs: Low $\chi_{\rm eff}$ mergers and predictions
  for LIGO}},}\ }\href {\doibase 10.1093/mnras/staa740} {\bibfield  {journal}
  {\bibinfo  {journal} {Mon. Not. Roy. Astron. Soc.}\ }\textbf {\bibinfo
  {volume} {494}},\ \bibinfo {pages} {1203--1216} (\bibinfo {year} {2020})},\
  \Eprint {http://arxiv.org/abs/1907.04356} {arXiv:1907.04356 [astro-ph.HE]}
  \BibitemShut {NoStop}%
%%CITATION = ARXIV:1907.04356;%%
\bibitem [{\citenamefont {Mandel}\ and\ \citenamefont
  {O'Shaughnessy}(2010)}]{Mandel:2009nx}%
  \BibitemOpen
  \bibfield  {author} {\bibinfo {author} {\bibfnamefont {Ilya}\ \bibnamefont
  {Mandel}}\ and\ \bibinfo {author} {\bibfnamefont {Richard}\ \bibnamefont
  {O'Shaughnessy}},\ }\bibfield  {title} {\enquote {\bibinfo {title} {{Compact
  Binary Coalescences in the Band of Ground-based Gravitational-Wave
  Detectors}},}\ }\bibfield  {booktitle} {\emph {\bibinfo {booktitle}
  {{Numerical relativity and data analysis. Proceedings, 3rd Annual Meeting,
  NRDA 2009, Potsdam, Germany, July 6-9, 2009}}},\ }\href {\doibase
  10.1088/0264-9381/27/11/114007} {\bibfield  {journal} {\bibinfo  {journal}
  {Class. Quant. Grav.}\ }\textbf {\bibinfo {volume} {27}},\ \bibinfo {pages}
  {114007} (\bibinfo {year} {2010})},\ \Eprint {http://arxiv.org/abs/0912.1074}
  {arXiv:0912.1074 [astro-ph.HE]} \BibitemShut {NoStop}%
%%CITATION = ARXIV:0912.1074;%%
\bibitem [{\citenamefont {Coleman~Miller}\ and\ \citenamefont
  {Krolik}(2013)}]{Miller:2013gya}%
  \BibitemOpen
  \bibfield  {author} {\bibinfo {author} {\bibfnamefont {M.}~\bibnamefont
  {Coleman~Miller}}\ and\ \bibinfo {author} {\bibfnamefont {Julian~H.}\
  \bibnamefont {Krolik}},\ }\bibfield  {title} {\enquote {\bibinfo {title}
  {{Alignment of supermassive black hole binary orbits and spins}},}\ }\href
  {\doibase 10.1088/0004-637X/774/1/43} {\bibfield  {journal} {\bibinfo
  {journal} {Astrophys. J.}\ }\textbf {\bibinfo {volume} {774}},\ \bibinfo
  {pages} {43} (\bibinfo {year} {2013})},\ \Eprint
  {http://arxiv.org/abs/1307.6569} {arXiv:1307.6569 [astro-ph.HE]} \BibitemShut
  {NoStop}%
%%CITATION = ARXIV:1307.6569;%%
\bibitem [{\citenamefont {Sesana}\ \emph {et~al.}(2014)\citenamefont {Sesana},
  \citenamefont {Barausse}, \citenamefont {Dotti},\ and\ \citenamefont
  {Rossi}}]{Sesana:2014bea}%
  \BibitemOpen
  \bibfield  {author} {\bibinfo {author} {\bibfnamefont {A.}~\bibnamefont
  {Sesana}}, \bibinfo {author} {\bibfnamefont {E.}~\bibnamefont {Barausse}},
  \bibinfo {author} {\bibfnamefont {M.}~\bibnamefont {Dotti}}, \ and\ \bibinfo
  {author} {\bibfnamefont {E.~M.}\ \bibnamefont {Rossi}},\ }\bibfield  {title}
  {\enquote {\bibinfo {title} {{Linking the spin evolution of massive black
  holes to galaxy kinematics}},}\ }\href {\doibase 10.1088/0004-637X/794/2/104}
  {\bibfield  {journal} {\bibinfo  {journal} {Astrophys. J.}\ }\textbf
  {\bibinfo {volume} {794}},\ \bibinfo {pages} {104} (\bibinfo {year}
  {2014})},\ \Eprint {http://arxiv.org/abs/1402.7088} {arXiv:1402.7088
  [astro-ph.CO]} \BibitemShut {NoStop}%
%%CITATION = ARXIV:1402.7088;%%
\bibitem [{\citenamefont {Gerosa}\ \emph
  {et~al.}(2015{\natexlab{a}})\citenamefont {Gerosa}, \citenamefont {Veronesi},
  \citenamefont {Lodato},\ and\ \citenamefont {Rosotti}}]{Gerosa:2015xya}%
  \BibitemOpen
  \bibfield  {author} {\bibinfo {author} {\bibfnamefont {Davide}\ \bibnamefont
  {Gerosa}}, \bibinfo {author} {\bibfnamefont {Benedetta}\ \bibnamefont
  {Veronesi}}, \bibinfo {author} {\bibfnamefont {Giuseppe}\ \bibnamefont
  {Lodato}}, \ and\ \bibinfo {author} {\bibfnamefont {Giovanni}\ \bibnamefont
  {Rosotti}},\ }\bibfield  {title} {\enquote {\bibinfo {title} {{Spin alignment
  and differential accretion in merging black hole binaries}},}\ }\href
  {\doibase 10.1093/mnras/stv1214} {\bibfield  {journal} {\bibinfo  {journal}
  {Mon. Not. Roy. Astron. Soc.}\ }\textbf {\bibinfo {volume} {451}},\ \bibinfo
  {pages} {3941--3954} (\bibinfo {year} {2015}{\natexlab{a}})},\ \Eprint
  {http://arxiv.org/abs/1503.06807} {arXiv:1503.06807 [astro-ph.GA]}
  \BibitemShut {NoStop}%
%%CITATION = ARXIV:1503.06807;%%
\bibitem [{\citenamefont {Gerosa}\ \emph
  {et~al.}(2015{\natexlab{b}})\citenamefont {Gerosa}, \citenamefont {Kesden},
  \citenamefont {O'Shaughnessy}, \citenamefont {Klein}, \citenamefont {Berti},
  \citenamefont {Sperhake},\ and\ \citenamefont
  {Trifir{\`o}}}]{Gerosa:2015hba}%
  \BibitemOpen
  \bibfield  {author} {\bibinfo {author} {\bibfnamefont {Davide}\ \bibnamefont
  {Gerosa}}, \bibinfo {author} {\bibfnamefont {Michael}\ \bibnamefont
  {Kesden}}, \bibinfo {author} {\bibfnamefont {Richard}\ \bibnamefont
  {O'Shaughnessy}}, \bibinfo {author} {\bibfnamefont {Antoine}\ \bibnamefont
  {Klein}}, \bibinfo {author} {\bibfnamefont {Emanuele}\ \bibnamefont {Berti}},
  \bibinfo {author} {\bibfnamefont {Ulrich}\ \bibnamefont {Sperhake}}, \ and\
  \bibinfo {author} {\bibfnamefont {Daniele}\ \bibnamefont {Trifir{\`o}}},\
  }\bibfield  {title} {\enquote {\bibinfo {title} {{Precessional instability in
  binary black holes with aligned spins}},}\ }\href {\doibase
  10.1103/PhysRevLett.115.141102} {\bibfield  {journal} {\bibinfo  {journal}
  {Phys. Rev. Lett.}\ }\textbf {\bibinfo {volume} {115}},\ \bibinfo {pages}
  {141102} (\bibinfo {year} {2015}{\natexlab{b}})},\ \Eprint
  {http://arxiv.org/abs/1506.09116} {arXiv:1506.09116 [gr-qc]} \BibitemShut
  {NoStop}%
\bibitem [{\citenamefont {Lousto}\ and\ \citenamefont
  {Healy}(2016)}]{Lousto:2016nlp}%
  \BibitemOpen
  \bibfield  {author} {\bibinfo {author} {\bibfnamefont {Carlos~O.}\
  \bibnamefont {Lousto}}\ and\ \bibinfo {author} {\bibfnamefont {James}\
  \bibnamefont {Healy}},\ }\bibfield  {title} {\enquote {\bibinfo {title}
  {{Unstable flip-flopping spinning binary black holes}},}\ }\href {\doibase
  10.1103/PhysRevD.93.124074} {\bibfield  {journal} {\bibinfo  {journal} {Phys.
  Rev. D}\ }\textbf {\bibinfo {volume} {93}},\ \bibinfo {pages} {124074}
  (\bibinfo {year} {2016})},\ \Eprint {http://arxiv.org/abs/1601.05086}
  {arXiv:1601.05086 [gr-qc]} \BibitemShut {NoStop}%
\bibitem [{\citenamefont {Mould}\ and\ \citenamefont
  {Gerosa}(2020)}]{Mould:2020cgc}%
  \BibitemOpen
  \bibfield  {author} {\bibinfo {author} {\bibfnamefont {Matthew}\ \bibnamefont
  {Mould}}\ and\ \bibinfo {author} {\bibfnamefont {Davide}\ \bibnamefont
  {Gerosa}},\ }\bibfield  {title} {\enquote {\bibinfo {title} {{Endpoint of the
  up-down instability in precessing binary black holes}},}\ }\href {\doibase
  10.1103/PhysRevD.101.124037} {\bibfield  {journal} {\bibinfo  {journal}
  {Phys. Rev. D}\ }\textbf {\bibinfo {volume} {101}},\ \bibinfo {pages}
  {124037} (\bibinfo {year} {2020})},\ \Eprint
  {http://arxiv.org/abs/2003.02281} {arXiv:2003.02281 [gr-qc]} \BibitemShut
  {NoStop}%
\bibitem [{\citenamefont {Bellovary}\ \emph {et~al.}(2016)\citenamefont
  {Bellovary}, \citenamefont {Mac~Low}, \citenamefont {McKernan},\ and\
  \citenamefont {Ford}}]{Bellovary:2015ifg}%
  \BibitemOpen
  \bibfield  {author} {\bibinfo {author} {\bibfnamefont {Jillian~M.}\
  \bibnamefont {Bellovary}}, \bibinfo {author} {\bibfnamefont {Mordecai-Mark}\
  \bibnamefont {Mac~Low}}, \bibinfo {author} {\bibfnamefont {Barry}\
  \bibnamefont {McKernan}}, \ and\ \bibinfo {author} {\bibfnamefont
  {K.~E.~Saavik}\ \bibnamefont {Ford}},\ }\bibfield  {title} {\enquote
  {\bibinfo {title} {{Migration Traps in Disks Around Supermassive Black
  Holes}},}\ }\href {\doibase 10.3847/2041-8205/819/2/L17} {\bibfield
  {journal} {\bibinfo  {journal} {Astrophys. J. Lett.}\ }\textbf {\bibinfo
  {volume} {819}},\ \bibinfo {pages} {L17} (\bibinfo {year} {2016})},\ \Eprint
  {http://arxiv.org/abs/1511.00005} {arXiv:1511.00005 [astro-ph.GA]}
  \BibitemShut {NoStop}%
\bibitem [{\citenamefont {Bartos}\ \emph {et~al.}(2017)\citenamefont {Bartos},
  \citenamefont {Kocsis}, \citenamefont {Haiman},\ and\ \citenamefont
  {M\'arka}}]{Bartos:2016dgn}%
  \BibitemOpen
  \bibfield  {author} {\bibinfo {author} {\bibfnamefont {Imre}\ \bibnamefont
  {Bartos}}, \bibinfo {author} {\bibfnamefont {Bence}\ \bibnamefont {Kocsis}},
  \bibinfo {author} {\bibfnamefont {Zolt}\ \bibnamefont {Haiman}}, \ and\
  \bibinfo {author} {\bibfnamefont {Szabolcs}\ \bibnamefont {M\'arka}},\
  }\bibfield  {title} {\enquote {\bibinfo {title} {{Rapid and Bright
  Stellar-mass Binary Black Hole Mergers in Active Galactic Nuclei}},}\ }\href
  {\doibase 10.3847/1538-4357/835/2/165} {\bibfield  {journal} {\bibinfo
  {journal} {Astrophys. J.}\ }\textbf {\bibinfo {volume} {835}},\ \bibinfo
  {pages} {165} (\bibinfo {year} {2017})},\ \Eprint
  {http://arxiv.org/abs/1602.03831} {arXiv:1602.03831 [astro-ph.HE]}
  \BibitemShut {NoStop}%
\bibitem [{\citenamefont {Stone}\ \emph {et~al.}(2017)\citenamefont {Stone},
  \citenamefont {Metzger},\ and\ \citenamefont {Haiman}}]{Stone:2016wzz}%
  \BibitemOpen
  \bibfield  {author} {\bibinfo {author} {\bibfnamefont {Nicholas~C.}\
  \bibnamefont {Stone}}, \bibinfo {author} {\bibfnamefont {Brian~D.}\
  \bibnamefont {Metzger}}, \ and\ \bibinfo {author} {\bibfnamefont {Zolt\'an}\
  \bibnamefont {Haiman}},\ }\bibfield  {title} {\enquote {\bibinfo {title}
  {{Assisted inspirals of stellar mass black holes embedded in AGN discs:
  solving the \textquoteleft{}final au problem\textquoteright{}}},}\ }\href
  {\doibase 10.1093/mnras/stw2260} {\bibfield  {journal} {\bibinfo  {journal}
  {Mon. Not. Roy. Astron. Soc.}\ }\textbf {\bibinfo {volume} {464}},\ \bibinfo
  {pages} {946--954} (\bibinfo {year} {2017})},\ \Eprint
  {http://arxiv.org/abs/1602.04226} {arXiv:1602.04226 [astro-ph.GA]}
  \BibitemShut {NoStop}%
\bibitem [{\citenamefont {Mckernan}\ \emph {et~al.}(2018)\citenamefont
  {Mckernan} \emph {et~al.}}]{McKernan:2017umu}%
  \BibitemOpen
  \bibfield  {author} {\bibinfo {author} {\bibfnamefont {B.}~\bibnamefont
  {Mckernan}} \emph {et~al.},\ }\bibfield  {title} {\enquote {\bibinfo {title}
  {{Constraining Stellar-mass Black Hole Mergers in AGN Disks Detectable with
  LIGO}},}\ }\href {\doibase 10.3847/1538-4357/aadae5} {\bibfield  {journal}
  {\bibinfo  {journal} {Astrophys. J.}\ }\textbf {\bibinfo {volume} {866}},\
  \bibinfo {pages} {66} (\bibinfo {year} {2018})},\ \Eprint
  {http://arxiv.org/abs/1702.07818} {arXiv:1702.07818 [astro-ph.HE]}
  \BibitemShut {NoStop}%
\bibitem [{\citenamefont {Yang}\ \emph
  {et~al.}(2019{\natexlab{b}})\citenamefont {Yang} \emph
  {et~al.}}]{Yang:2019cbr}%
  \BibitemOpen
  \bibfield  {author} {\bibinfo {author} {\bibfnamefont {Yang}\ \bibnamefont
  {Yang}} \emph {et~al.},\ }\bibfield  {title} {\enquote {\bibinfo {title}
  {{Hierarchical Black Hole Mergers in Active Galactic Nuclei}},}\ }\href
  {\doibase 10.1103/PhysRevLett.123.181101} {\bibfield  {journal} {\bibinfo
  {journal} {Phys. Rev. Lett.}\ }\textbf {\bibinfo {volume} {123}},\ \bibinfo
  {pages} {181101} (\bibinfo {year} {2019}{\natexlab{b}})},\ \Eprint
  {http://arxiv.org/abs/1906.09281} {arXiv:1906.09281 [astro-ph.HE]}
  \BibitemShut {NoStop}%
%%CITATION = ARXIV:1906.09281;%%
\bibitem [{\citenamefont {Kesden}\ \emph {et~al.}(2015)\citenamefont {Kesden},
  \citenamefont {Gerosa}, \citenamefont {O'Shaughnessy}, \citenamefont
  {Berti},\ and\ \citenamefont {Sperhake}}]{Kesden:2014sla}%
  \BibitemOpen
  \bibfield  {author} {\bibinfo {author} {\bibfnamefont {Michael}\ \bibnamefont
  {Kesden}}, \bibinfo {author} {\bibfnamefont {Davide}\ \bibnamefont {Gerosa}},
  \bibinfo {author} {\bibfnamefont {Richard}\ \bibnamefont {O'Shaughnessy}},
  \bibinfo {author} {\bibfnamefont {Emanuele}\ \bibnamefont {Berti}}, \ and\
  \bibinfo {author} {\bibfnamefont {Ulrich}\ \bibnamefont {Sperhake}},\
  }\bibfield  {title} {\enquote {\bibinfo {title} {{Effective potentials and
  morphological transitions for binary black-hole spin precession}},}\ }\href
  {\doibase 10.1103/PhysRevLett.114.081103} {\bibfield  {journal} {\bibinfo
  {journal} {Phys. Rev. Lett.}\ }\textbf {\bibinfo {volume} {114}},\ \bibinfo
  {pages} {081103} (\bibinfo {year} {2015})},\ \Eprint
  {http://arxiv.org/abs/1411.0674} {arXiv:1411.0674 [gr-qc]} \BibitemShut
  {NoStop}%
%%CITATION = ARXIV:1411.0674;%%
\bibitem [{\citenamefont {Gerosa}\ \emph
  {et~al.}(2015{\natexlab{c}})\citenamefont {Gerosa}, \citenamefont {Kesden},
  \citenamefont {Sperhake}, \citenamefont {Berti},\ and\ \citenamefont
  {O'Shaughnessy}}]{Gerosa:2015tea}%
  \BibitemOpen
  \bibfield  {author} {\bibinfo {author} {\bibfnamefont {Davide}\ \bibnamefont
  {Gerosa}}, \bibinfo {author} {\bibfnamefont {Michael}\ \bibnamefont
  {Kesden}}, \bibinfo {author} {\bibfnamefont {Ulrich}\ \bibnamefont
  {Sperhake}}, \bibinfo {author} {\bibfnamefont {Emanuele}\ \bibnamefont
  {Berti}}, \ and\ \bibinfo {author} {\bibfnamefont {Richard}\ \bibnamefont
  {O'Shaughnessy}},\ }\bibfield  {title} {\enquote {\bibinfo {title}
  {{Multi-timescale analysis of phase transitions in precessing black-hole
  binaries}},}\ }\href {\doibase 10.1103/PhysRevD.92.064016} {\bibfield
  {journal} {\bibinfo  {journal} {Phys. Rev. D}\ }\textbf {\bibinfo {volume}
  {92}},\ \bibinfo {pages} {064016} (\bibinfo {year} {2015}{\natexlab{c}})},\
  \Eprint {http://arxiv.org/abs/1506.03492} {arXiv:1506.03492 [gr-qc]}
  \BibitemShut {NoStop}%
\bibitem [{SXS()}]{SXSWebsite}%
  \BibitemOpen
  \href@noop {} {\enquote {\bibinfo {title} {Simulating e{X}treme
  {S}pacetimes},}\ }\bibinfo {note}
  {\href{http://www.black-holes.org/}{black-holes.org}}\BibitemShut {NoStop}%
\bibitem [{\citenamefont {Boyle}\ \emph {et~al.}(2019)\citenamefont {Boyle}
  \emph {et~al.}}]{Boyle:2019kee}%
  \BibitemOpen
  \bibfield  {author} {\bibinfo {author} {\bibfnamefont {Michael}\ \bibnamefont
  {Boyle}} \emph {et~al.},\ }\bibfield  {title} {\enquote {\bibinfo {title}
  {{The SXS Collaboration catalog of binary black hole simulations}},}\ }\href
  {\doibase 10.1088/1361-6382/ab34e2} {\bibfield  {journal} {\bibinfo
  {journal} {Class. Quant. Grav.}\ }\textbf {\bibinfo {volume} {36}},\ \bibinfo
  {pages} {195006} (\bibinfo {year} {2019})},\ \Eprint
  {http://arxiv.org/abs/1904.04831} {arXiv:1904.04831 [gr-qc]} \BibitemShut
  {NoStop}%
%%CITATION = ARXIV:1904.04831;%%
\bibitem [{\citenamefont {Szil\'agyi}\ \emph {et~al.}(2015)\citenamefont
  {Szil\'agyi}, \citenamefont {Blackman}, \citenamefont {Buonanno},
  \citenamefont {Taracchini}, \citenamefont {Pfeiffer}, \citenamefont {Scheel},
  \citenamefont {Chu}, \citenamefont {Kidder},\ and\ \citenamefont
  {Pan}}]{Szilagyi:2015rwa}%
  \BibitemOpen
  \bibfield  {author} {\bibinfo {author} {\bibfnamefont {B\'ela}\ \bibnamefont
  {Szil\'agyi}}, \bibinfo {author} {\bibfnamefont {Jonathan}\ \bibnamefont
  {Blackman}}, \bibinfo {author} {\bibfnamefont {Alessandra}\ \bibnamefont
  {Buonanno}}, \bibinfo {author} {\bibfnamefont {Andrea}\ \bibnamefont
  {Taracchini}}, \bibinfo {author} {\bibfnamefont {Harald~P.}\ \bibnamefont
  {Pfeiffer}}, \bibinfo {author} {\bibfnamefont {Mark~A.}\ \bibnamefont
  {Scheel}}, \bibinfo {author} {\bibfnamefont {Tony}\ \bibnamefont {Chu}},
  \bibinfo {author} {\bibfnamefont {Lawrence~E.}\ \bibnamefont {Kidder}}, \
  and\ \bibinfo {author} {\bibfnamefont {Yi}~\bibnamefont {Pan}},\ }\bibfield
  {title} {\enquote {\bibinfo {title} {{Approaching the Post-Newtonian Regime
  with Numerical Relativity: A Compact-Object Binary Simulation Spanning 350
  Gravitational-Wave Cycles}},}\ }\href {\doibase
  10.1103/PhysRevLett.115.031102} {\bibfield  {journal} {\bibinfo  {journal}
  {Phys. Rev. Lett.}\ }\textbf {\bibinfo {volume} {115}},\ \bibinfo {pages}
  {031102} (\bibinfo {year} {2015})},\ \Eprint
  {http://arxiv.org/abs/1502.04953} {arXiv:1502.04953 [gr-qc]} \BibitemShut
  {NoStop}%
\bibitem [{\citenamefont {Boyle}\ and\ \citenamefont
  {Mroue}(2009)}]{Boyle:2009vi}%
  \BibitemOpen
  \bibfield  {author} {\bibinfo {author} {\bibfnamefont {Michael}\ \bibnamefont
  {Boyle}}\ and\ \bibinfo {author} {\bibfnamefont {Abdul~H.}\ \bibnamefont
  {Mroue}},\ }\bibfield  {title} {\enquote {\bibinfo {title} {{Extrapolating
  gravitational-wave data from numerical simulations}},}\ }\href {\doibase
  10.1103/PhysRevD.80.124045} {\bibfield  {journal} {\bibinfo  {journal} {Phys.
  Rev.}\ }\textbf {\bibinfo {volume} {D80}},\ \bibinfo {pages} {124045}
  (\bibinfo {year} {2009})},\ \Eprint {http://arxiv.org/abs/0905.3177}
  {arXiv:0905.3177 [gr-qc]} \BibitemShut {NoStop}%
%%CITATION = ARXIV:0905.3177;%%
\bibitem [{\citenamefont {Boyle}(2016)}]{Boyle:2015nqa}%
  \BibitemOpen
  \bibfield  {author} {\bibinfo {author} {\bibfnamefont {Michael}\ \bibnamefont
  {Boyle}},\ }\bibfield  {title} {\enquote {\bibinfo {title} {{Transformations
  of asymptotic gravitational-wave data}},}\ }\href {\doibase
  10.1103/PhysRevD.93.084031} {\bibfield  {journal} {\bibinfo  {journal} {Phys.
  Rev.}\ }\textbf {\bibinfo {volume} {D93}},\ \bibinfo {pages} {084031}
  (\bibinfo {year} {2016})},\ \Eprint {http://arxiv.org/abs/1509.00862}
  {arXiv:1509.00862 [gr-qc]} \BibitemShut {NoStop}%
%%CITATION = ARXIV:1509.00862;%%
\bibitem [{\citenamefont {Varma}\ \emph
  {et~al.}(2019{\natexlab{a}})\citenamefont {Varma}, \citenamefont {Field},
  \citenamefont {Scheel}, \citenamefont {Blackman}, \citenamefont {Gerosa},
  \citenamefont {Stein}, \citenamefont {Kidder},\ and\ \citenamefont
  {Pfeiffer}}]{Varma:2019csw}%
  \BibitemOpen
  \bibfield  {author} {\bibinfo {author} {\bibfnamefont {Vijay}\ \bibnamefont
  {Varma}}, \bibinfo {author} {\bibfnamefont {Scott~E.}\ \bibnamefont {Field}},
  \bibinfo {author} {\bibfnamefont {Mark~A.}\ \bibnamefont {Scheel}}, \bibinfo
  {author} {\bibfnamefont {Jonathan}\ \bibnamefont {Blackman}}, \bibinfo
  {author} {\bibfnamefont {Davide}\ \bibnamefont {Gerosa}}, \bibinfo {author}
  {\bibfnamefont {Leo~C.}\ \bibnamefont {Stein}}, \bibinfo {author}
  {\bibfnamefont {Lawrence~E.}\ \bibnamefont {Kidder}}, \ and\ \bibinfo
  {author} {\bibfnamefont {Harald~P.}\ \bibnamefont {Pfeiffer}},\ }\bibfield
  {title} {\enquote {\bibinfo {title} {{Surrogate models for precessing binary
  black hole simulations with unequal masses}},}\ }\href {\doibase
  10.1103/PhysRevResearch.1.033015} {\bibfield  {journal} {\bibinfo  {journal}
  {Phys. Rev. Research.}\ }\textbf {\bibinfo {volume} {1}},\ \bibinfo {pages}
  {033015} (\bibinfo {year} {2019}{\natexlab{a}})},\ \Eprint
  {http://arxiv.org/abs/1905.09300} {arXiv:1905.09300 [gr-qc]} \BibitemShut
  {NoStop}%
%%CITATION = ARXIV:1905.09300;%%
\bibitem [{\citenamefont {Blackman}\ \emph {et~al.}(2017)\citenamefont
  {Blackman}, \citenamefont {Field}, \citenamefont {Scheel}, \citenamefont
  {Galley}, \citenamefont {Ott}, \citenamefont {Boyle}, \citenamefont {Kidder},
  \citenamefont {Pfeiffer},\ and\ \citenamefont
  {Szil{\'a}gyi}}]{Blackman:2017pcm}%
  \BibitemOpen
  \bibfield  {author} {\bibinfo {author} {\bibfnamefont {Jonathan}\
  \bibnamefont {Blackman}}, \bibinfo {author} {\bibfnamefont {Scott~E.}\
  \bibnamefont {Field}}, \bibinfo {author} {\bibfnamefont {Mark~A.}\
  \bibnamefont {Scheel}}, \bibinfo {author} {\bibfnamefont {Chad~R.}\
  \bibnamefont {Galley}}, \bibinfo {author} {\bibfnamefont {Christian~D.}\
  \bibnamefont {Ott}}, \bibinfo {author} {\bibfnamefont {Michael}\ \bibnamefont
  {Boyle}}, \bibinfo {author} {\bibfnamefont {Lawrence~E.}\ \bibnamefont
  {Kidder}}, \bibinfo {author} {\bibfnamefont {Harald~P.}\ \bibnamefont
  {Pfeiffer}}, \ and\ \bibinfo {author} {\bibfnamefont {B{\'e}la}\ \bibnamefont
  {Szil{\'a}gyi}},\ }\bibfield  {title} {\enquote {\bibinfo {title} {{Numerical
  relativity waveform surrogate model for generically precessing binary black
  hole mergers}},}\ }\href {\doibase 10.1103/PhysRevD.96.024058} {\bibfield
  {journal} {\bibinfo  {journal} {Phys. Rev.}\ }\textbf {\bibinfo {volume}
  {D96}},\ \bibinfo {pages} {024058} (\bibinfo {year} {2017})},\ \Eprint
  {http://arxiv.org/abs/1705.07089} {arXiv:1705.07089 [gr-qc]} \BibitemShut
  {NoStop}%
%%CITATION = ARXIV:1705.07089;%%
\bibitem [{\citenamefont {Ossokine}\ \emph {et~al.}(2015)\citenamefont
  {Ossokine}, \citenamefont {Boyle}, \citenamefont {Kidder}, \citenamefont
  {Pfeiffer}, \citenamefont {Scheel},\ and\ \citenamefont
  {Szil{\'a}gyi}}]{Ossokine:2015vda}%
  \BibitemOpen
  \bibfield  {author} {\bibinfo {author} {\bibfnamefont {Serguei}\ \bibnamefont
  {Ossokine}}, \bibinfo {author} {\bibfnamefont {Michael}\ \bibnamefont
  {Boyle}}, \bibinfo {author} {\bibfnamefont {Lawrence~E.}\ \bibnamefont
  {Kidder}}, \bibinfo {author} {\bibfnamefont {Harald~P.}\ \bibnamefont
  {Pfeiffer}}, \bibinfo {author} {\bibfnamefont {Mark~A.}\ \bibnamefont
  {Scheel}}, \ and\ \bibinfo {author} {\bibfnamefont {B{\'e}la}\ \bibnamefont
  {Szil{\'a}gyi}},\ }\bibfield  {title} {\enquote {\bibinfo {title} {{Comparing
  Post-Newtonian and Numerical-Relativity Precession Dynamics}},}\ }\href
  {\doibase 10.1103/PhysRevD.92.104028} {\bibfield  {journal} {\bibinfo
  {journal} {Phys. Rev.}\ }\textbf {\bibinfo {volume} {D92}},\ \bibinfo {pages}
  {104028} (\bibinfo {year} {2015})},\ \Eprint
  {http://arxiv.org/abs/1502.01747} {arXiv:1502.01747 [gr-qc]} \BibitemShut
  {NoStop}%
%%CITATION = ARXIV:1502.01747;%%
\bibitem [{\citenamefont {Zhao}\ \emph {et~al.}(2017)\citenamefont {Zhao},
  \citenamefont {Kesden},\ and\ \citenamefont {Gerosa}}]{Zhao:2017tro}%
  \BibitemOpen
  \bibfield  {author} {\bibinfo {author} {\bibfnamefont {Xinyu}\ \bibnamefont
  {Zhao}}, \bibinfo {author} {\bibfnamefont {Michael}\ \bibnamefont {Kesden}},
  \ and\ \bibinfo {author} {\bibfnamefont {Davide}\ \bibnamefont {Gerosa}},\
  }\bibfield  {title} {\enquote {\bibinfo {title} {{Nutational resonances,
  transitional precession, and precession-averaged evolution in binary
  black-hole systems}},}\ }\href {\doibase 10.1103/PhysRevD.96.024007}
  {\bibfield  {journal} {\bibinfo  {journal} {Phys. Rev.}\ }\textbf {\bibinfo
  {volume} {D96}},\ \bibinfo {pages} {024007} (\bibinfo {year} {2017})},\
  \Eprint {http://arxiv.org/abs/1705.02369} {arXiv:1705.02369 [gr-qc]}
  \BibitemShut {NoStop}%
%%CITATION = ARXIV:1705.02369;%%
\bibitem [{\citenamefont {Buonanno}\ \emph {et~al.}(2009)\citenamefont
  {Buonanno}, \citenamefont {Iyer}, \citenamefont {Ochsner}, \citenamefont
  {Pan},\ and\ \citenamefont {Sathyaprakash}}]{Buonanno:2009zt}%
  \BibitemOpen
  \bibfield  {author} {\bibinfo {author} {\bibfnamefont {Alessandra}\
  \bibnamefont {Buonanno}}, \bibinfo {author} {\bibfnamefont {Bala}\
  \bibnamefont {Iyer}}, \bibinfo {author} {\bibfnamefont {Evan}\ \bibnamefont
  {Ochsner}}, \bibinfo {author} {\bibfnamefont {Yi}~\bibnamefont {Pan}}, \ and\
  \bibinfo {author} {\bibfnamefont {B.S.}\ \bibnamefont {Sathyaprakash}},\
  }\bibfield  {title} {\enquote {\bibinfo {title} {{Comparison of
  post-Newtonian templates for compact binary inspiral signals in
  gravitational-wave detectors}},}\ }\href {\doibase
  10.1103/PhysRevD.80.084043} {\bibfield  {journal} {\bibinfo  {journal} {Phys.
  Rev. D}\ }\textbf {\bibinfo {volume} {80}},\ \bibinfo {pages} {084043}
  (\bibinfo {year} {2009})},\ \Eprint {http://arxiv.org/abs/0907.0700}
  {arXiv:0907.0700 [gr-qc]} \BibitemShut {NoStop}%
\bibitem [{\citenamefont {Isoyama}\ \emph {et~al.}(2020)\citenamefont
  {Isoyama}, \citenamefont {Sturani},\ and\ \citenamefont
  {Nakano}}]{Isoyama:2020lls}%
  \BibitemOpen
  \bibfield  {author} {\bibinfo {author} {\bibfnamefont {Soichiro}\
  \bibnamefont {Isoyama}}, \bibinfo {author} {\bibfnamefont {Riccardo}\
  \bibnamefont {Sturani}}, \ and\ \bibinfo {author} {\bibfnamefont {Hiroyuki}\
  \bibnamefont {Nakano}},\ }\bibfield  {title} {\enquote {\bibinfo {title}
  {{Post-Newtonian templates for gravitational waves from compact binary
  inspirals}},}\ }\href@noop {} {\  (\bibinfo {year} {2020})},\ \Eprint
  {http://arxiv.org/abs/2012.01350} {arXiv:2012.01350 [gr-qc]} \BibitemShut
  {NoStop}%
\bibitem [{\citenamefont {{LIGO and Virgo Collaborations}}(2018)}]{lalsuite}%
  \BibitemOpen
  \bibfield  {author} {\bibinfo {author} {\bibnamefont {{LIGO and Virgo
  Collaborations}}},\ }\href {\doibase 10.7935/GT1W-FZ16} {\enquote {\bibinfo
  {title} {{LIGO} {A}lgorithm {L}ibrary - {LALS}uite},}\ } (\bibinfo {year}
  {2018})\BibitemShut {NoStop}%
\bibitem [{\citenamefont {Buonanno}\ \emph {et~al.}(2003)\citenamefont
  {Buonanno}, \citenamefont {Chen},\ and\ \citenamefont
  {Vallisneri}}]{Buonanno:2002fy}%
  \BibitemOpen
  \bibfield  {author} {\bibinfo {author} {\bibfnamefont {Alessandra}\
  \bibnamefont {Buonanno}}, \bibinfo {author} {\bibfnamefont {Yan-bei}\
  \bibnamefont {Chen}}, \ and\ \bibinfo {author} {\bibfnamefont {Michele}\
  \bibnamefont {Vallisneri}},\ }\bibfield  {title} {\enquote {\bibinfo {title}
  {{Detecting gravitational waves from precessing binaries of spinning compact
  objects: Adiabatic limit}},}\ }\href {\doibase 10.1103/PhysRevD.67.104025}
  {\bibfield  {journal} {\bibinfo  {journal} {Phys. Rev. D}\ }\textbf {\bibinfo
  {volume} {67}},\ \bibinfo {pages} {104025} (\bibinfo {year} {2003})},\
  \bibinfo {note} {[Erratum: Phys.Rev.D 74, 029904 (2006)]},\ \Eprint
  {http://arxiv.org/abs/gr-qc/0211087} {arXiv:gr-qc/0211087} \BibitemShut
  {NoStop}%
\bibitem [{\citenamefont {Ajith}(2011)}]{Ajith:2011ec}%
  \BibitemOpen
  \bibfield  {author} {\bibinfo {author} {\bibfnamefont {P.}~\bibnamefont
  {Ajith}},\ }\bibfield  {title} {\enquote {\bibinfo {title} {{Addressing the
  spin question in gravitational-wave searches: Waveform templates for
  inspiralling compact binaries with nonprecessing spins}},}\ }\href {\doibase
  10.1103/PhysRevD.84.084037} {\bibfield  {journal} {\bibinfo  {journal} {Phys.
  Rev. D}\ }\textbf {\bibinfo {volume} {84}},\ \bibinfo {pages} {084037}
  (\bibinfo {year} {2011})},\ \Eprint {http://arxiv.org/abs/1107.1267}
  {arXiv:1107.1267 [gr-qc]} \BibitemShut {NoStop}%
%%CITATION = ARXIV:1107.1267;%%
\bibitem [{\citenamefont {Ossokine}\ \emph {et~al.}(2020)\citenamefont
  {Ossokine} \emph {et~al.}}]{Ossokine:2020kjp}%
  \BibitemOpen
  \bibfield  {author} {\bibinfo {author} {\bibfnamefont {Serguei}\ \bibnamefont
  {Ossokine}} \emph {et~al.},\ }\bibfield  {title} {\enquote {\bibinfo {title}
  {{Multipolar Effective-One-Body Waveforms for Precessing Binary Black Holes:
  Construction and Validation}},}\ }\href {\doibase
  10.1103/PhysRevD.102.044055} {\bibfield  {journal} {\bibinfo  {journal}
  {Phys. Rev. D}\ }\textbf {\bibinfo {volume} {102}},\ \bibinfo {pages}
  {044055} (\bibinfo {year} {2020})},\ \Eprint
  {http://arxiv.org/abs/2004.09442} {arXiv:2004.09442 [gr-qc]} \BibitemShut
  {NoStop}%
\bibitem [{\citenamefont {Owen}\ \emph {et~al.}(2019)\citenamefont {Owen},
  \citenamefont {Fox}, \citenamefont {Freiberg},\ and\ \citenamefont
  {Jacques}}]{Owen:2017yaj}%
  \BibitemOpen
  \bibfield  {author} {\bibinfo {author} {\bibfnamefont {Robert}\ \bibnamefont
  {Owen}}, \bibinfo {author} {\bibfnamefont {Alex~S.}\ \bibnamefont {Fox}},
  \bibinfo {author} {\bibfnamefont {John~A.}\ \bibnamefont {Freiberg}}, \ and\
  \bibinfo {author} {\bibfnamefont {Terrence~Pierre}\ \bibnamefont {Jacques}},\
  }\bibfield  {title} {\enquote {\bibinfo {title} {{Black Hole Spin Axis in
  Numerical Relativity}},}\ }\href {\doibase 10.1103/PhysRevD.99.084031}
  {\bibfield  {journal} {\bibinfo  {journal} {Phys. Rev. D}\ }\textbf {\bibinfo
  {volume} {99}},\ \bibinfo {pages} {084031} (\bibinfo {year} {2019})},\
  \Eprint {http://arxiv.org/abs/1708.07325} {arXiv:1708.07325 [gr-qc]}
  \BibitemShut {NoStop}%
\bibitem [{\citenamefont {Varma}\ \emph
  {et~al.}(2019{\natexlab{b}})\citenamefont {Varma}, \citenamefont {Field},
  \citenamefont {Scheel}, \citenamefont {Blackman}, \citenamefont {Kidder},\
  and\ \citenamefont {Pfeiffer}}]{Varma:2018mmi}%
  \BibitemOpen
  \bibfield  {author} {\bibinfo {author} {\bibfnamefont {Vijay}\ \bibnamefont
  {Varma}}, \bibinfo {author} {\bibfnamefont {Scott~E.}\ \bibnamefont {Field}},
  \bibinfo {author} {\bibfnamefont {Mark~A.}\ \bibnamefont {Scheel}}, \bibinfo
  {author} {\bibfnamefont {Jonathan}\ \bibnamefont {Blackman}}, \bibinfo
  {author} {\bibfnamefont {Lawrence~E.}\ \bibnamefont {Kidder}}, \ and\
  \bibinfo {author} {\bibfnamefont {Harald~P.}\ \bibnamefont {Pfeiffer}},\
  }\bibfield  {title} {\enquote {\bibinfo {title} {{Surrogate model of
  hybridized numerical relativity binary black hole waveforms}},}\ }\href
  {\doibase 10.1103/PhysRevD.99.064045} {\bibfield  {journal} {\bibinfo
  {journal} {Phys. Rev.}\ }\textbf {\bibinfo {volume} {D99}},\ \bibinfo {pages}
  {064045} (\bibinfo {year} {2019}{\natexlab{b}})},\ \Eprint
  {http://arxiv.org/abs/1812.07865} {arXiv:1812.07865 [gr-qc]} \BibitemShut
  {NoStop}%
%%CITATION = ARXIV:1812.07865;%%
\bibitem [{\citenamefont {Stanzione}\ \emph {et~al.}(2020)\citenamefont
  {Stanzione}, \citenamefont {West}, \citenamefont {Evans}, \citenamefont
  {Minyard}, \citenamefont {Ghattas},\ and\ \citenamefont {Panda}}]{Frontera}%
  \BibitemOpen
  \bibfield  {author} {\bibinfo {author} {\bibfnamefont {Dan}\ \bibnamefont
  {Stanzione}}, \bibinfo {author} {\bibfnamefont {John}\ \bibnamefont {West}},
  \bibinfo {author} {\bibfnamefont {R.~Todd}\ \bibnamefont {Evans}}, \bibinfo
  {author} {\bibfnamefont {Tommy}\ \bibnamefont {Minyard}}, \bibinfo {author}
  {\bibfnamefont {Omar}\ \bibnamefont {Ghattas}}, \ and\ \bibinfo {author}
  {\bibfnamefont {Dhabaleswar~K.}\ \bibnamefont {Panda}},\ }\bibfield  {title}
  {\enquote {\bibinfo {title} {Frontera: The evolution of leadership computing
  at the national science foundation},}\ }in\ \href {\doibase
  10.1145/3311790.3396656} {\emph {\bibinfo {booktitle} {Practice and
  Experience in Advanced Research Computing}}},\ \bibinfo {series and number}
  {PEARC '20}\ (\bibinfo  {publisher} {Association for Computing Machinery},\
  \bibinfo {address} {New York, NY, USA},\ \bibinfo {year} {2020})\ pp.\
  \bibinfo {pages} {106--111}\BibitemShut {NoStop}%
\end{thebibliography}%

\end{document}